%% file: main.tex
\documentclass[conference]{IEEEtran}
\usepackage{cite}
\usepackage{amsmath,amssymb,amsfonts}
\usepackage{algorithmic}
\usepackage{graphicx}
\usepackage{textcomp}
\usepackage{xcolor}
\usepackage{fancyhdr}
\usepackage[hyphens]{url}

\def\BibTeX{{\rm B\kern-.05em{\sc i\kern-.025em b}\kern-.08em
    T\kern-.1667em\lower.7ex\hbox{E}\kern-.125emX}}

\usepackage{threeparttable}
\usepackage[caption=false,font=footnotesize]{subfig}

\usepackage[normalem]{ulem}
\usepackage{ifthen}
\usepackage{dblfloatfix}

\newboolean{publicversion}
\setboolean{publicversion}{false}
\ifthenelse{\boolean{publicversion}}{
  \newcommand{\grumbler}[2]{}
  \newcommand{\inplace}[2]{}
}{
  \newcommand{\grumbler}[2]{\marginpar{{\bf #1:}{\textcolor{red}{#2}}}}
  \newcommand{\inplace}[2]{{\bf #1:} {\textcolor{red}{#2}}}
  
}
\title{Beyond Application End-Point Results: Quantifying Statistical Robustness of MCMC Accelerators} 
\author{
\IEEEauthorblockN{Xiangyu Zhang, Ramin Bashizade, Yicheng Wang, Cheng Lyu, Sayan Mukherjee, Alvin R. Lebeck} \\
\IEEEauthorblockA{\textit{Duke University}\\
\textit{\{xiangyu.zhang, ramin.bashizade, yicheng.wang, cheng.lu3\}@duke.edu, sayan@stat.duke.edu, alvy@cs.duke.edu}
}
}
\begin{document}
\maketitle
\pagestyle{plain}

%%%%%% -- PAPER CONTENT STARTS-- %%%%%%%%
\begin{abstract}
    \input{Abstract}
\end{abstract}

\input{Intro}
\input{Background}

\input{Pillars}
\input{Results}
\input{DesignSpace}

\input{LimFuture}
\input{RelatedWork}
\input{Conclusion}

\section*{Acknowledgements}
This project is supported in part by Intel, the Semiconductor Research Corporation and the National Science Foundation (CNS-1616947). 

%%%%%%% -- PAPER CONTENT ENDS -- %%%%%%%%

%%%%%%%%% -- BIB STYLE AND FILE -- %%%%%%%%
\bibliographystyle{IEEEtranS}
\bibliography{ref}
%%%%%%%%%%%%%%%%%%%%%%%%%%%%%%%%%%%%

\end{document}

%% file: Abstract.tex
Statistical machine learning often uses probabilistic algorithms, such as Markov Chain Monte Carlo (MCMC), to solve a wide range of problems.
Probabilistic computations, often considered too slow on conventional processors, can be accelerated with specialized hardware by exploiting parallelism and optimizing the design using various approximation techniques. Current methodologies for evaluating correctness of probabilistic accelerators are often incomplete, mostly focusing only on end-point result quality (``accuracy").
It is important for hardware designers and domain experts to look beyond end-point ``accuracy" and be aware of the hardware optimizations impact on other statistical properties. 

This work takes a first step towards defining metrics and a methodology for quantitatively evaluating correctness of probabilistic accelerators beyond end-point result quality. We propose three pillars of statistical robustness: 1) sampling quality, 2) convergence diagnostic, and 3) goodness of fit. 
We apply our framework to a representative MCMC accelerator and surface design issues that cannot be exposed using only application end-point result quality. Applying the framework to guide design space exploration shows that statistical robustness comparable to floating-point software can be achieved
by slightly increasing the bit representation, without floating-point hardware requirements. 

%% file: Intro.tex
\section{Introduction}

Statistical machine learning, like other methods in artificial intelligence, has become an important workload for computing systems. Such workloads often utilize probabilistic algorithms, such as Markov Chain Monte Carlo (MCMC) methods, which enable the potential to provide generalized frameworks to solve a wide range of problems. As alternatives to Deep Neural Networks, these algorithms provide easier access to interpreting why a given result is obtained through their model transparency and statistical properties. 
Many specialized accelerators are proposed to address the sampling inefficiency of probabilistic algorithms \cite{Wang:isca2016,mansinghka2014building,liu2015,mahajan2016tabla,mingas2017particle,ko2017,acmc2019,cai2018vibnn}, by utilizing approximation techniques to improve the hardware efficiency, such as reducing bit representation, truncating small values to zero, or simplifying the random number generator. 

Understanding the influence of these approximations on the application results is crucial to meet the quality requirement in real applications. A hardware accelerator should provide correct execution of target algorithms. A common approach to evaluating correctness is to compare the end-point result quality (``accuracy") against accurately-measured or hand-labeled ground-truth data using community-standard benchmarks and metrics: the hardware execution is considered to be correct if it provides comparable ``accuracy'' to the software-only implementations that do not have these approximations. 
However, in the domain of probabilistic computing/algorithms, correctness is defined by more than the end-point result of executing the algorithm, and includes additional statistical properties that convey uncertainty and interpretability about the end-point result. End-point ``accuracy" is necessary but not sufficient to claim correctness. Current methodologies for evaluating probabilistic accelerators are often incomplete or adhoc in evaluating correctness, focusing only on end-point ``accuracy" or limited statistical properties.  Failure to adequately account for domain-defined correctness can have adverse or catastrophic outcomes, such as a surgeon failing to completely remove a tumor due to incorrect uncertainty in a segmented image \cite{cheng2016joint,mcclure2019}.
Furthermore, end-point ``accuracy'' may not always be possible since ground-truth data is not always available.
It is important for hardware designers and domain experts to look beyond end-point ``accuracy'' and be aware of the impact of optimizations on other statistical properties. Therefore, \textit{a probabilistic architecture should provide some measure (or guarantee) of statistical robustness.}

This work takes a first step towards defining metrics and a methodology for quantitatively evaluating correctness of probabilistic accelerators beyond end-point result quality. 
We propose three pillars of statistical robustness: 1) \textit{sampling quality}, 2) \textit{convergence diagnostic}, and 3) \textit{goodness of fit} \textbf{(Contribution 1)}. 
Each pillar has at least one quantitative empirical metric, does not require ground-truth data, and collectively these pillars enable comparison of specialized hardware to 64-bit floating-point (FP64) software. We expose several challenges with naively applying existing popular metrics for our purposes, including: high dimensionality of the target applications, and random variables with zero variance. Therefore, we modify the existing methodologies for sampling quality and convergence diagnostic, and propose a new metric for convergence diagnostic \textbf{(Contribution 2)}. We call on domain experts to develop metrics with stronger theoretical foundations to account for common hardware optimizations. Below is a summary of each pillar. 

\textit{Pillar I) Sampling Quality.} The intrinsic nature of MCMC methods creates dependency between samples. A sufficient number of independent samples are needed to converge and produce high-quality results. We use \textit{Effective Sample Size} (ESS) to measure the number of independent samples drawn from an MCMC run, and report the arithmetic mean as a scalar metric. Low ESS indicates that more iterations may be required to generate sufficient independent samples. 

\textit{Pillar II) Convergence Diagnostic.} The total running time of an MCMC run is determined by when it converges. Convergence can be measured by Gelman-Rubin's $\hat{R}$ \cite{brooksgelman98}, but this metric is undefined for variables with zero variance. Therefore, we propose a process to determine convergence that accounts for zero variance and a new metric---\textit{convergence percentage}---based on 
$\hat{R}$, to measure the total percentage of converged results. Low convergence percentage indicates that more iterations are required for the model to converge. 

\textit{Pillar III) Goodness of Fit.} In the absence of ground-truth data (labeled data), it is important to understand the differences between the baseline FP64 and hardware end-point results to evaluate the overall quality of the hardware. 
We provide two ``goodness of fit" approaches: 1) Root Mean Squared Error (RMSE) on application specific data relative to a software reference, and 2) Jensen-Shannon Divergence (JSD) to evaluate all possible data inputs and provide worst-case distribution divergence. 

As a case study, we demonstrate the framework in a representative probabilistic accelerator~\cite{zhang2018isca} and show that 1) end-point accuracy alone is insufficient, particularly for predicting outcome for previously unseen inputs, and 2) that FP64 is insufficient as ground truth since in some cases more limited precision can produce more accurate end-point results based on labeled data \textbf{(Contribution 3)}.
The analysis shows that the accelerator achieves the same application end-point result quality as FP64 software, confirming the previous work. However, we show the accelerator differs with FP64 in ESS and convergence percentage, and reveal that applications need to run $2\times$ more iterations on the accelerator to achieve the same statistical robustness as FP64, reducing the accelerator's effective speedup.  
We also explore the design space of the above accelerator to expose the trade-offs between statistical robustness and area/power \textbf{(Contribution 4)}. We find that considerable improvement in statistical robustness, comparable to the FP64 software, can be achieved
by slightly increasing the bit precision from 4 to 6 and removing an approximation technique, with only  1.20$\times$ area and 1.10$\times$ power overhead.

The remainder of this paper is organized as follows. Sec. \ref{sec:background} provide the necessary background and motivation for this work.  The detailed description of three pillars is in Sec. \ref{sec:pillars}. Sec. \ref{sec:analysis} describes the analysis of statistical robustness on a representative accelerator and we perform design space exploration using the three pillars in Sec. \ref{sec:dse}. Sec. \ref{sec:future} discusses limitations and future work, related work is presented in Sec. \ref{sec:related}, and Sec. \ref{sec:conclusion} concludes the paper. 

%% file: Background.tex
\section{Background and Motivation}\label{sec:background}

\subsection{Probabilistic Algorithms}\label{sec:prob_algs}

Probabilistic (stochastic) algorithms are a powerful approach used in many application (e.g., image analysis \cite{geman1984stochastic}, robotics, natural language processing \cite{finkel2005incorporating}, global health \cite{hamra2013markov}, and wireless communications \cite{hedstrom2017mimo}). Probabilistic algorithms solve problems by randomly inferring the parameters in the probabilistic models to explain observed data, which create opportunities to provide generalized frameworks and are the only practical approach to solve certain problems, such as high-dimensional inference. 
Compared with deep neural networks, 
probabilistic models are ``conceptually simple, compositional, and interpretable" \cite{ghahramani2015probabilistic}. 
Bayesian Inference is an important framework in probabilistic models, which updates the probability estimate for a hypothesis by combining information from prior beliefs and observed data. Suppose $X$ is the latent random variable of interest. The goal is to retrieve the posterior distribution $p(X|D)$ given the prior beliefs of $X$ and the observed data $D$ using Bayes' theorem: $p(X|D) \propto p(D|X)p(X)$, where $p(X)$ is the prior distribution of $X$, and $p(D|X)$ is the likelihood of observing $D$ given a certain value of $X$. 
Both $X$ and $D$ can be scalars or multidimensional vectors. As the dimension of $D$ and $X$ increases, analytically or numerically deriving the exact result of a posterior distribution becomes complicated and intractable. 
One approach to break ``the curse of dimensionality" is Markov Chain Monte Carlo (MCMC), 
methods that solve the problems by iteratively generating samples on random variables
and eventually converge to a result regardless of the initial stage.

\begin{figure}
        \centering
  	\includegraphics[width=0.48\textwidth,keepaspectratio]{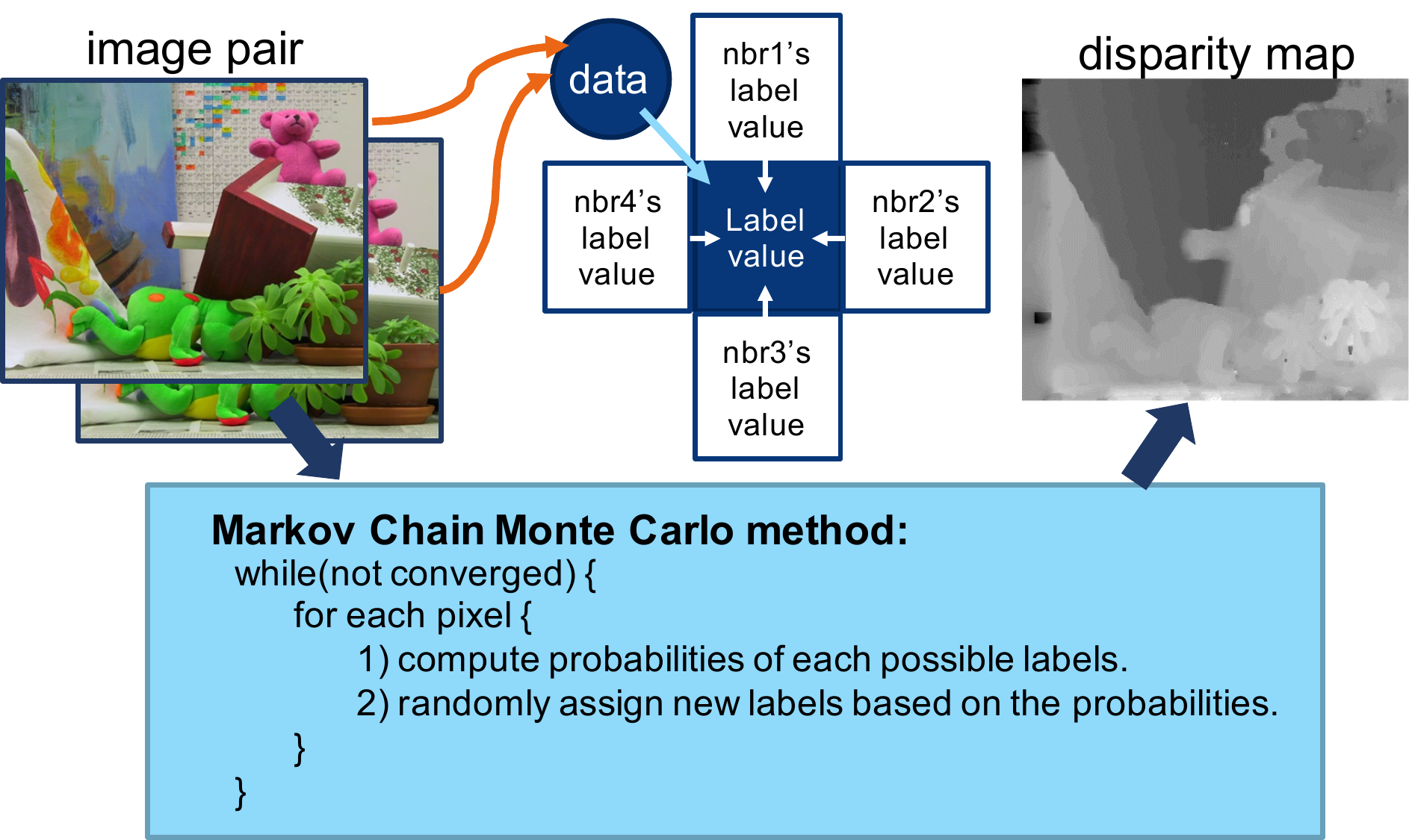} 
        \caption{Stereo vision using Markov Chain Monte Carlo (Markov Random Field Gibbs Sampling). Note that Sampling is performed in the inner loop.}
        \label{fig:mcmc}
\end{figure}

Fig. \ref{fig:mcmc} shows an example of using an MCMC method, Markov Random Field (MRF) Gibbs Sampling, in stereo vision, demonstrated by previous work \cite{Barnard1989}. Stereo vision reconstructs the depth information of objects in an image pair by matching the corresponding pixels that represent the same objects. 
The results are presented in a disparity map, indicating the depth of the corresponding objects in the image (lighter is closer). As shown in Fig. \ref{fig:mcmc}, the MCMC method iterates the image pixels by considering the disparity of each pixel as a random variable. 
For each pixel, it evaluates probabilities of each possible label (disparity outcome) and draws a sample as the output label. Each probability is determined by the neighboring pixels label values and the initial pixel data values of the image pair, defined by First-Order MRF graphical model \cite{geman1984stochastic}. The outer loop (a.k.a. iteration) iterates on the whole image until convergence.

MCMC methods rely on efficiently generating samples from a parameterized distribution, which involves step 1) efficiently computing the parameters of the distribution to sample from based on observed data, and step 2) efficiently generating samples from the parameterized distribution. Unfortunately, as described in previous work \cite{Wang:isca2016}, sampling overhead can be notably high: step 2) alone takes hundreds of CPU cycles for a simple distribution. One approach to reduce the sampling overhead is to use approximation techniques in algorithms \cite{neal2011mcmc,welling2011bayesian,martino2016orthogonal}.
Deterministic methods, such as Expectation Propagation and Variational Bayesian, are alternatives to MCMC. Although these methods are often more efficient in the applied cases, domain experts prefer conceptually straightforward, mathematically simple, yet accurate methods.

\subsection{Specialized Probabilistic Accelerators}\label{sec:spec_acc}
 
Meeting the needs of domain experts may be achieved by accelerating sampling through hardware specialization. 
Previous efforts seek to efficiently generate specific types of distributions using FPGAs~\cite{thomas2009using,bachir2008new}, specialized circuits~\cite{chakrapani2006ultra}, specialized architectures are proposed to accelerate specific algorithms and models, such as a Stochastic Transition Circuit~\cite{mansinghka2014building}, dedicated MCMC models~\cite{liu2015,mingas2017particle,Ko2018}, Bayesian Neural Network~\cite{cai2018vibnn},
Stochastic Gradient Descent~\cite{mahajan2016tabla}, an ASIC accelerator for Bayesian Networks~\cite{khan2016hardware},  CMOS-hybrid MRF Gibbs Sampling Unit~\cite{Wang:isca2016,zhang2018isca}, and workflows to compile probabilistic programming 
language to hardware accelerators in Chisel code~\cite{acmc2019}.
Many of these accelerators use various forms of approximation (e.g., limited bit precision, pseudo random number generators, etc.) to reduce area/power, allowing more individual units on a single chip and thus improving overall performance.  

Understanding the influence of these hardware optimizations on correctness is a critical aspect of any design process.  An intuitive approach is to evaluate the end-point (final) result quality (``accuracy") for applications using community-standard benchmarks and metrics. 
However, end-point ``accuracy" is necessary but not sufficient to claim correctness: 1) given the uncertainty of input data, the observed end-point result quality has no indication of ``accuracy" for unseen data, and thus just making statements on the observed ``accuracy" is not enough, 2) many applications look beyond the end-point ``accuracy" and consider uncertainty quantification, 
and 3) measuring ``accuracy" may not always be possible as ground-truth data is not always accessible and thus no comparison to an end-point result is available. Probabilistic algorithm domain experts frequently use statistical properties to evaluate the confidence on the unseen data and to help uncertainty quantification (e.g., tumor resection \cite{cheng2016joint,mcclure2019}). 
\textit{A probabilistic architecture should provide some measure (or guarantee) of statistical robustness.}

This paper represents a first step toward articulating a set of metrics and methodology for quantifying the statistical robustness of probabilistic accelerators. Sec.~\ref{sec:pillars} presents our proposed metrics and we use an accelerator from Zhang, et al.~\cite{zhang2018isca} (described below) as a case study to demonstrate how to analyze an existing accelerator and to perform design space exploration.

\subsection{A Representative Probabilistic Accelerator}
\label{sec:SPU}

\begin{figure}
        \centering
  	\includegraphics[width=0.48\textwidth,keepaspectratio, trim=8mm 0mm 0mm 0mm]{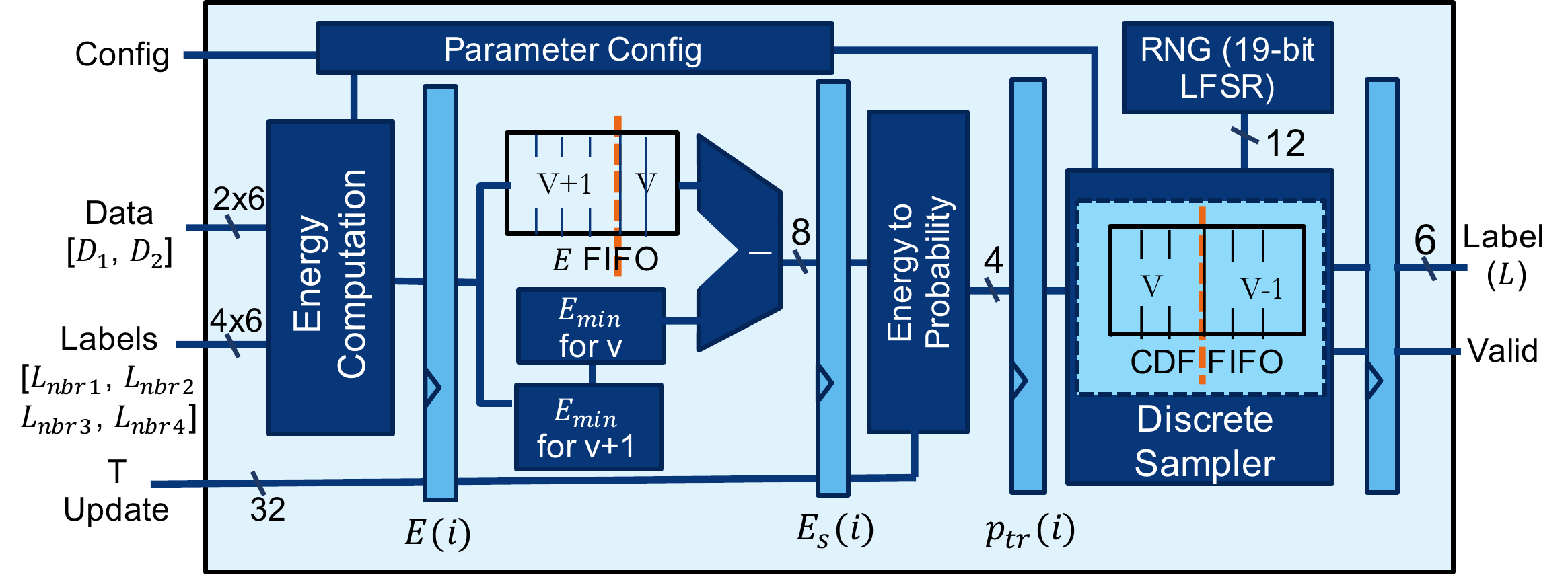} 
        \caption{The SPU pipeline derived from RSU-G \cite{zhang2018isca}. The sampling stage is replaced with a CMOS discrete sampler.}
        \label{fig:spu}
\end{figure}

As a case study, we consider the Gibbs Sampling accelerator design described by Zhang, et al.~\cite{zhang2018isca} implemented entirely in CMOS using pseudo random number generation instead of molecular optical devices (cf. \cite{zhang2018isca} Sec. IV.C).  Fig.~\ref{fig:spu} shows the baseline pipeline design, which we call a Stochastic Processing Unit (SPU). It is divided into four main stages with two internal decoupling FIFOs and an inverse transform method is used for the discrete sampler.

Given the data and neighbor labels, the first stage computes the total energy of a possible label $E(i)$ (Eq. 
\ref{eq:energy}) each cycle, where $\alpha$ and $\beta$ are application parameters. The energy $E(i)$ is then dynamically scaled using subtraction to maximize the dynamic 
range (Eq. \ref{eq:dy_scaling}).  Both $E(i)$ and $E_s(i)$ are 8-bit unsigned integers. 
In the third stage, the scaled energy $E_s(i)$ is converted to a scaled probability represented in 4-bit unsigned 
integer. The original probability is computed by $exp(-E_s(i)/T)$ which is represented as a real number between [0,1] using floating-point in software, where $T$ is a fixed parameter per outer iteration. However, the probability is 
scaled using Eq. \ref{eq:untr_prob} and truncated using Eq. \ref{eq:tr_prob} to match the unsigned integer 
representation, where $P_{bits}=4$ is the number of bits in the scaled probability output $p_{tr}(i)$. Additionally, 
probability truncation drives all scaled probabilities that are less than one to zero and $2^n$ approximation rounds 
all scaled probabilities down to the nearest $2^n$ integer value (Eq. \ref{eq:tr_prob}). The value of $p_{tr}(i)$ 
can be pre-computed and stored in a look-up table (LUT). The values in the LUT need updates if $T$ changes between iterations. The final stage of SPU generates a discrete sample per variable based on the probabilities of all possible label values $\{p_{tr}(0), p_{tr}(1), ...\}$ using the least 12-bits of a 19-bit LFSR to implement the inverse transform sampling.

\begin{equation} \label{eq:energy}
E(i) = \alpha E_{singleton}(i) + \beta \sum{E_{neighborhood}}
\end{equation}
\begin{equation} \label{eq:dy_scaling}
E_s(i) = E(i) - E_{min}
\end{equation}
\begin{equation} \label{eq:untr_prob}
p_s(i) = (2^{P_{bits}}-1)\times exp(-E_s(i)/T)
\end{equation}
\begin{equation} \label{eq:tr_prob}
p_{tr}(i) = \lfloor 2^{\lfloor \log_2 p_s(i) \rfloor} \rfloor 
\end{equation}

The SPU supports two operating modes: 1) pure-sampling and 2) optimization (simulated annealing).   Pure-sampling iteratively generates Gibbs samples using constant temperature $T$, where $T$ is considered a parameter obtained during model training. When converged, the estimated distribution of a random variable (e.g., distribution of possible disparities in a pixel) can be obtained by collecting the latest N samples. An exact result can be obtained from the mode of the estimated distribution, the most frequent label. The optimization mode uses simulated annealing to converge to an exact result faster by strategically decreasing the temperature $T$~\cite{geman1984stochastic}.  $T$ is initially high so that every label has similar probability of being selected. %in the beginning. 
As $T$ decreases, labels with the lowest energy are likely to be selected, eventually leading to convergence. The optimization mode requires fewer iterations than pure-sampling, but cannot provide an estimated distribution. The previous work~\cite{zhang2018isca} evaluates only the optimization mode. 

We implement the SPU in Verilog and use QuestaSim simulation to evaluate the end-point result quality of the same three applications assessed in the previous work~\cite{zhang2018isca}: image segmentation, motion estimation, and stereo vision. We use community-standard metrics (e.g., variation of information in image segmentation \cite{martin2001imseg,Yang2008unsupervised}, end-point error in motion estimation \cite{Baker2011motion}, and bad-pixel percentage in stereo vision \cite{scharstein2002stereo}). 
Tab. \ref{table:spu_prelim_results} shows the result quality comparison between FP64 software and the SPU. Each result is collected by a single run per dataset in simulated annealing mode.  We validate that
the SPU with a simple 19-bit LFSR as its random number generator (RNG) achieves the same result quality as the software. Image segmentation results indicate the same conclusion and are omitted for brevity. We also obtain similar high quality applicatoin results on an Intel Arria 10 FPGA prototype. 
Despite these results we are left with the question: \textit{What do the results in Tab. \ref{table:spu_prelim_results} indicate about accelerator statistical robustness}? The short answer is \textit{nothing}.  The following sections present our initial efforts toward providing a better answer.

\begin{scriptsize}
\begin{table}
  \centering
  \begin{threeparttable}
  \caption{SPU result quality from one run per dataset. }
  \label{table:spu_prelim_results}
  \begin{tabular}{|l|l|l|l|l|l|}
    \hline
    \multicolumn{3}{|c|}{\textbf{Motion estimation\tnote{1}}} & \multicolumn{3}{|c|}{\textbf{Stereo vision\tnote{2}}} \\
    \hline
    \hline
    Dataset & Software & SPU & Dataset & Software & SPU \\
    \hline
    \textit{dimetrodon} & 0.600 & 0.611 & \textit{art} & 26.8\% & 27.7\% \\
    \hline
    \textit{rubberwhale} & 0.371 & 0.376 & \textit{poster} & 12.3\% & 11.0\% \\
    \hline
    \textit{venus} & 0.460 & 0.449 & \textit{teddy} & 26.9\% & 27.8\% \\
    \hline
  \end{tabular}
  \begin{tablenotes}
      \item [1] Metric is end-point error. Lower is better.
      \item [2] Metric is bad-pixel percentage. Lower is better.
  \end{tablenotes}
  \end{threeparttable}
\end{table}
\end{scriptsize}

%% file: Pillars.tex
\section{Three Pillars of Statistical Robustness}
\label{sec:pillars}

To identify appropriate measures of hardware statistical robustness, we draw on known techniques utilized by domain experts to evaluate their models and algorithms.  Ideally, we could formally prove bounds on relevant metrics~\cite{ge2018simulated}.
Unfortunately, some hardware optimizations (e.g., truncation to zero) make formal proofs extremely difficult or impossible. A provable architecture introduces more complicated hardware. Therefore, we rely on existing empirical diagnostic tests for MCMC techniques, that are based on foundations in statistics, to establish three pillars for assessing a probabilistic accelerator's statistical robustness: 1) sampling quality~\cite{thompson2010comparison}, 2) convergence~\cite{cowles1996markov}, and 3) goodness of fit. Each pillar has at least one quantitative measure, and provides insight to application users and 
to hardware designers.  Collectively these pillars help in understanding/explaining end-point results, and can
indicate the performance of the MCMC execution, such as how many iterations are required to converge. 
Note that the statistical robustness is jointly affected by the algorithm and hardware architecture. Therefore, we compare hardware results with FP64 software as the baseline to extract the impact of hardware optimizations. The remainder of this section presents our proposed three pillars for evaluating statistical robustness of an MCMC accelerator. 

\subsection{Pillar 1: Sampling Quality}
A sampling algorithm with perfect sampling quality generates independent samples. However, an MCMC sample is drawn based on the current values of random variables---the outcome of samples in the previous iteration. Recall in stereo vision, the disparity of a pixel depends on the current states of its neighborhood. This dependency creates correlations between samples which is nontrival until several subsequent samples are drawn, which can be represented as an autocorrelation time $\tau$. This implies that by generating $n$ samples from MCMC, only $n/\tau$ samples can be considered independent.  A sufficient number of independent samples are required to derive meaningful statistical measures (e.g., mean and variance). Note that the sample dependency is an intrinsic property of MCMC algorithms and exists even with a perfect random number generator and FP64 precision.

\textit{Effective Sample Size} (ESS) is commonly used as a sampling quality metric that represents how many independent samples are drawn among all the dependent samples. In general, higher ESS indicates the MCMC sampler is more efficient at generating independent samples. Unfortunately, there is not consensus on a single ESS definition~\cite{gong2016practical}. This paper uses the definition discussed by Kass et al.~\cite{kass1998markov} based on autocorrelation. Since closed form expressions for ESS are difficult we estimate ESS using known methods~\cite{thompson2010comparison,liu2015}.

We estimate ESS on a univariate random variable using Eq.~\ref{eq:ess}, where $n$ is the number of dependent MCMC samples (iterations) and $\rho(k)$ is the autocorrelation function of the sample sequence. We sum up the first $K$ contiguous lags where $\rho(k)+\rho(k+1) \geq 0$. Theoretically, $ESS=n$ provides the best sampling quality where all samples are independent; however, Eq. \ref{eq:ess} is an estimate of ESS, thus it is numerically possible that $ESS>n$. Furthermore, ESS has no definition when all collected samples have the same value (zero variance), which is possible in practice as shown in Sec. \ref{sec:analysis}. 
{%\small
\begin{equation}\label{eq:ess}
    ESS=n/(1+2\sum^{K}_{k=1}\rho(k))
\end{equation}
}
Many MCMC problems are high-dimensional (many random variables). For example, in stereo vision a 320$\times$320 input image has 102,400 dimensions. An ideal metric can report a scalar ESS value to account for all dimensions as a single random variable. While methods exist to report multivariate ESS \cite{vats2015multivariate}, to 
our knowledge they aren't practical in our case and they do not allow zero variance for any variable. 
In this paper, we calculate ESS per dimension (random variable/pixel) and report the arithmetic mean ESS. Sec. \ref{sec:analysis} describes how we handle zero variance cases, and we encourage domain experts to develop an evaluation method with stronger theoretical foundations to account for these zero variance cases. 

\textbf{Pillar Insight.} If ESS is low it may take more MCMC iterations to achieve an acceptable ESS.  If a hardware accelerator produces substantially lower ESS than software, the additional iterations may reduce its effective speedup.

\subsection{Pillar 2: Convergence Diagnostic}

An important question for an MCMC method is when to stop iterating, determined by when the MCMC is converged. Similar to effective sample size (ESS), the time to convergence is used to analyze algorithms and input data when using software even with FP64 and good random number generators. Multiple methods exist to measure the convergence. A comprehensive review is provided by Cowles et al. \cite{cowles1996markov}. We use Gelman-Rubin's $\hat{R}$ \cite{brooksgelman98}, a popular quantitative method provided by many statistical packages, to measure whether a univariate random variable (e.g., a pixel in stereo vision) is converged at a certain iteration. 

\begin{figure}
        \centering
  	\includegraphics[width=0.48\textwidth,keepaspectratio]{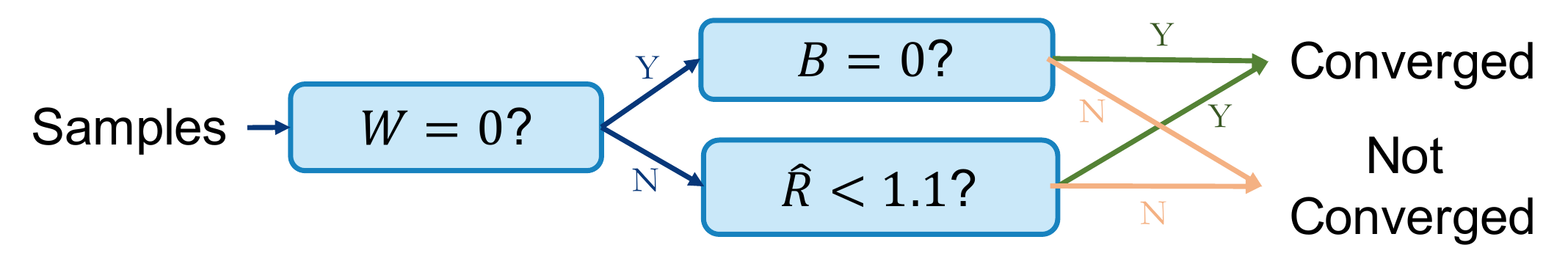} 
        \caption{Process to determine convergence of a univariate random variable.}
        \label{fig:conv_process}
\end{figure}

Gelman-Rubin's $\hat{R}$ (potential scale reduction factor) estimates the convergence by comparing the between-chain variance ($B$) and within-chain variance ($W$) across multiple independent runs on the same MCMC instance\footnote{Instance refers to the same input data, model and configuration parameters.}. Equations \ref{eq:sig_p} and \ref{eq:rhat} show the computation to obtain an $\hat{R}$ from $m$ independent MCMC runs, each with $n$ samples, where $\sigma^2_{+}$ is an estimate on the variance of random variable. As a rule of thumb \cite{gelman1992inference,brooksgelman98}, a univariate random variable is considered converged when $\hat{R}<1.1$. Typically larger $\hat{R}$ indicates that more iterations are needed to converge. 
Note that the $\hat{R}$ method requires a random value initialized from an overdispersed distribution. We meet this requirement by initializing random variables (i.e., initial labels of pixels) uniform-randomly. 
{
\begin{equation}\label{eq:sig_p}
    \hat{\sigma}^2_{+} = (n-1)/n\times W + B/n
\end{equation}
\begin{equation}\label{eq:rhat}
    \hat{R}^2=\frac{m+1}{m}\frac{\hat{\sigma}^2_{+}}{W}-\frac{n-1}{mn}
\end{equation}
}
 A scalar convergence diagnostic is preferred for multi-dimensional applications. Similar to ESS, handling high dimensions and the random variables with estimated zero variance ($W=0$) is challenging using existing methods  \cite{brooksgelman98,vats2018revisiting}. The original Gelman-Rubin's $\hat{R}$ metric has no definition at $W=0$. We consider a random variable converged when $B=0$ and $W=0$, which indicates all samples are the same value from different iterations and MCMC runs. A random variable is not considered converged when $B>0$ and $W=0$, which indicates samples are the same value within MCMC runs, but different across MCMC runs. Fig. \ref{fig:conv_process} shows our process to determine whether a univariate random variable is converged. We propose \textit{convergence percentage}, the percentage of converged univariate random variables, as our new metric.

\textbf{Pillar Insight.} Low convergence percentage indicates that more iterations are needed for the model to converge. If a hardware accelerator takes substantially more iterations to converge than software, the additional iterations may reduce its effective speedup.

\subsection{Pillar 3: Goodness of Fit}

Finally, understanding the ``goodness of fit"---the difference between end-point results produced by the software and by the hardware accelerator---is critical to evaluate the overall quality of the hardware accelerator. 
A straightforward approach is to compare the end-point result quality using community-standard benchmarks and metrics. However, ground truth data are not always available. 
We provide two ``goodness of fit" approaches: 1) using application specific data to measure how good the hardware results fit to a reference software result, and 2) using a distribution divergence measurement to evaluate all possible data inputs and provide worst-case divergence. 

\subsubsection{Application Data Analysis}
We are interested in how close/different the results are between the software and hardware. For example, the difference of disparities across the whole image given the same image inputs. Popular quantitative metrics for ``goodness of fit" include Root Mean Squared Error (RMSE) and coefficient of determination ($R^2$). 
RMSE measures the average squared difference between the result from a hardware MCMC run and a reference software run. The range of RMSE is from 0 to infinity, where lower is better. Zero RMSE means the hardware produces identical results to the software (i.e., perfectly fit). $R^2$ measures the portion of variance in the hardware results that can be explained by the software results. Typically, $R^2$ has a more intuitive range of [0,1] than RMSE. A higher $R^2$ value is preferred and $R^2=1$ means hardware and software have perfect fit.
However, variance of the software result is in the denominator of a negative term in $R^2$ formula. That means results can have a misleadingly bad $R^2$ value even with a good RMSE if variance of the software results is small. \textit{Since our target is to measure the result differences caused by hardware approximations, we choose using RMSE as our metric.} A detailed introduction to these statistics can be found elsewhere \cite{james2013introduction}. 

Due to the stochastic nature of MCMC methods, each MCMC run can have different end-point results for either software or hardware. To account for this variation, we compute RMSE for both hardware and software with respect to a reference software result from multiple MCMC runs. The reference software result is obtained using the mode of multiple software runs to minimize the result variation in a single software reference run.

\subsubsection{Data-independent Analysis}
Recall from Sec. \ref{sec:prob_algs} that step-one of sampling is computing the probability distribution to sample from. Hardware approximations in this step, such as limited precision and truncation, introduce divergence from the distribution obtained from FP64 software. 
Quantifying the distribution divergence of hardware $P_{hw}$ from software $P_{sw}$ provides insights on why the results are good (or bad) and gives the worst-case divergence in arbitrary data inputs. 
This can be measured using a popular divergence measure Kullback-Leibler (KL) divergence ($D_{KL}$). However, $D_{KL}$ value goes to infinity when any entry of $P_{hw}(i)$ is zero while $P_{sw}(i)$ is non-zero, which is likely to happen under the hardware technique of truncating small probabilities to zero, and thus cannot be directly applied in our study. 
We use Jensen-Shannon Divergence (JSD) \cite{lin1991divergence} instead as the divergence measurement. JSD, shown in Eq. \ref{eq:jsd}, is defined based on KL-devergence, where $M = (1/2) P_{sw}+ (1/2) P_{hw}$. Note that unlike KL-divergence, $D_{JS}$ is a symmetric method where $D_{JS}(P_{sw}||P_{hw}) = D_{JS}(P_{hw}||P_{sw})$. 

\begin{equation}\label{eq:jsd}
\begin{split}
    D_{JS}(P_{sw}||P_{hw})= \frac{1}{2} D_{KL}(P_{sw}||M) + \frac{1}{2} D_{KL}(P_{hw}||M) 
\end{split}
\end{equation}

 The goal of using JSD is to evaluate the distribution divergence caused by hardware approximations on arbitrary data inputs, which gives some insights on how 
 the hardware performs on unobserved data and provides worst-case scenarios. Unfortunately, evaluating JSD on arbitrary data inputs for a random variable with many possible labels, such as in stereo vision, is challenging in both analytical and empirical approaches given the complicated mathematical representation and the large parameter space. 
 In this work, we start from evaluating the JSD between software and hardware results with arbitrary data inputs when a random variable has two possible labels, such as in a foreground-background image segmentation. Analytical studies, as well as analysis on many-label cases, is our future work. 
 
\textbf{Pillar Insight.} Substantially worse RMSE or JSD results for a hardware accelerator means it is likely producing low quality application end-point results and more iterations or model/hardware design changes may be required.

%% file: Results.tex
\section{Analyzing Existing Hardware}
\label{sec:analysis}
 We apply the three pillars of statistical robustness on an existing hardware, the Stochastic Processing Unit (SPU) described in Sec.~\ref{sec:SPU}.

\subsection{Methodology}
In this work we consider a single SPU, as it is sufficient to explore the statistical robustness questions.  Development of an accelerator prototype with multiple SPUs is ongoing but is beyond the scope of this work. We primarily utilize MATLAB for both the FP64 software and for a functionally equivalent SPU simulator.  Importantly, we also have SPU implementations in Verilog, Chisel, and HLS all with verified results. The single-run result quality in Tab. \ref{table:spu_prelim_results} are from the Verilog implementation.

% To efficiently evaluate the statistical robustness of various hardware, we developed a MATLAB simulation and evaluation framework. The MATLAB implementations of 64-bit floating-point software applications and an SPU functionally equivalent simulator produce end-point results and records all sample traces throughout MCMC runs. The output results are sent to a self-developed evaluation package with some external codes \mike{cite if possible} for statistical robustness analysis. Theoretically, the evaluation package can support other MCMC accelerators with modifications on the data interface. The evaluation of JSD is a separate package that does not require application data. A MATLAB-C++ fusion code is also implemented in software stereo vision for finding appropriate application parameters faster. 

We choose stereo vision and motion estimation as our test applications. Motion estimation infers the motion vector of image pixels in a frame of a video with respect to its next sequential frame. The concept of applying MRF Gibbs Sampling on motion estimation is similar to stereo vision as described in Sec. \ref{sec:prob_algs}, except the output label is a 2D motion vector of a pixel, indicating where the pixel moves to in the next frame. Each disparity per pixel in stereo vision is treated as a random variable. Each 2D motion vector per pixel in motion estimation is considered as two random variables $x$ and $y$. 
We pick three datasets from Middlebury \cite{scharstein2002stereo,Baker2011motion} for each application, as in the previous work \cite{zhang2018isca}. We use FP64 runs to find the application parameters (e.g., $\alpha$ and $\beta$). Motion estimation has one set of parameters for all datasets, and stereo vision has two sets for all datasets. Some parameters can be optimized in a training process, which is beyond the scope of this paper. We also considered, but omit, image segmentation since it converges too fast (30 iterations for simulated annealing) to produce meaningful statistical measurements.

% We also explore image segmentation as a third application, but across 30 input data sets, we find that it converges in 30 iterations with simulated annealing and produces good result quality. We believe the number of samples per pixel is inadequate for statistically meaningful results, and thus we omit this application from our robustness analysis. 

Recall the SPU supports two operating modes: pure sampling that produces the full estimated distribution (sampling), and optimization using simulated-annealing (optimization) that converges quickly to an exact result.  For optimization, measuring Effective Sample Size (ESS) and Gelman Rubin's $\hat R$  is not conceptually meaningful, we evaluate sampling quality and convergence diagnostic for sampling only and goodness of fit for both modes.  
Parameter settings for each dataset are the same in sampling as in optimization, except for a different, fixed temperature. Our empirical results show that all datasets converge after 1,000 iterations for optimization and 3,000 for sampling, except for \textit{poster} in stereo vision which takes only 500 and 1,500 iterations, respectively. 

\subsection{Results Analysis}
\begin{figure}
        \centering
  	\includegraphics[width=0.48\textwidth,keepaspectratio]{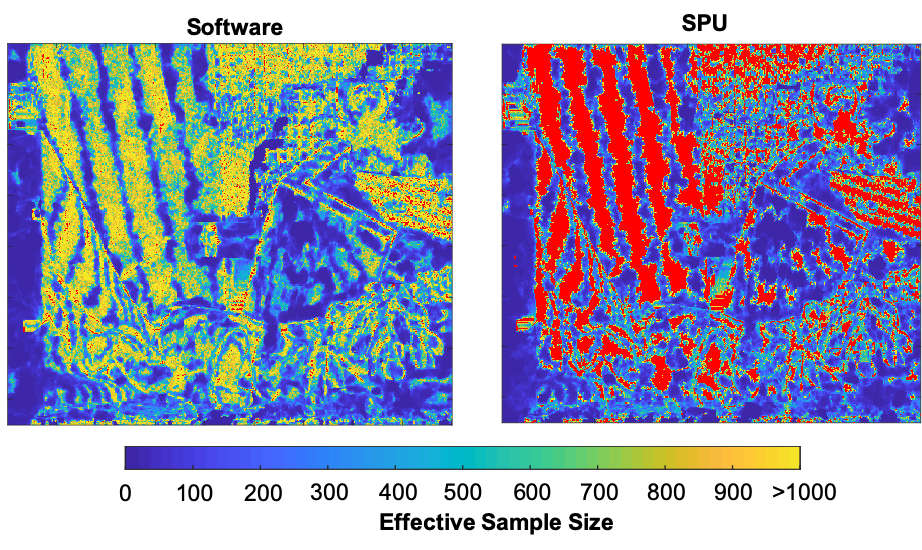} 
        \caption{ESS per random variable in stereo vision \textit{teddy}. Red regions correspond to zero variance. Red regions in the SPU overlap high ESS regions with small variances in the software.}
        \label{fig:ess_per_rv}
\end{figure}
\begin{figure}
\centering
	\captionsetup[subfigure]{width=0.20\textwidth}
	\subfloat[Stereo vision]{
		\includegraphics[width=0.23\textwidth,keepaspectratio,trim=5mm 3mm 0mm 3mm]{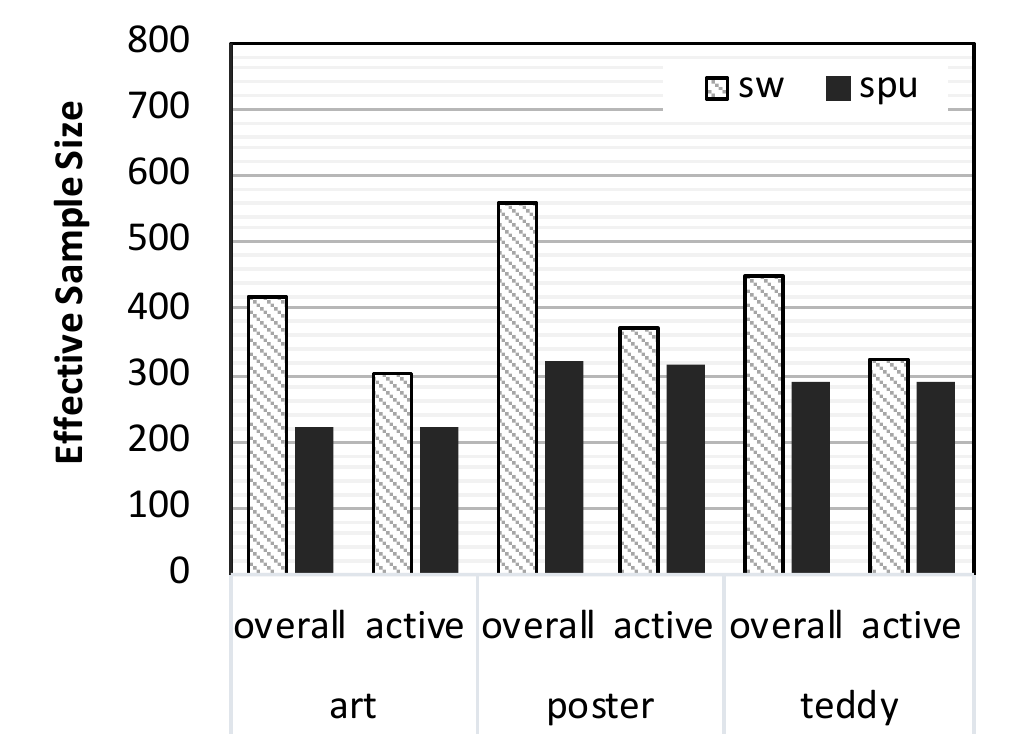}
    }
    \subfloat[Motion estimation]{
    	\includegraphics[width=0.23\textwidth,keepaspectratio,trim=5mm 8mm 0mm 3mm]{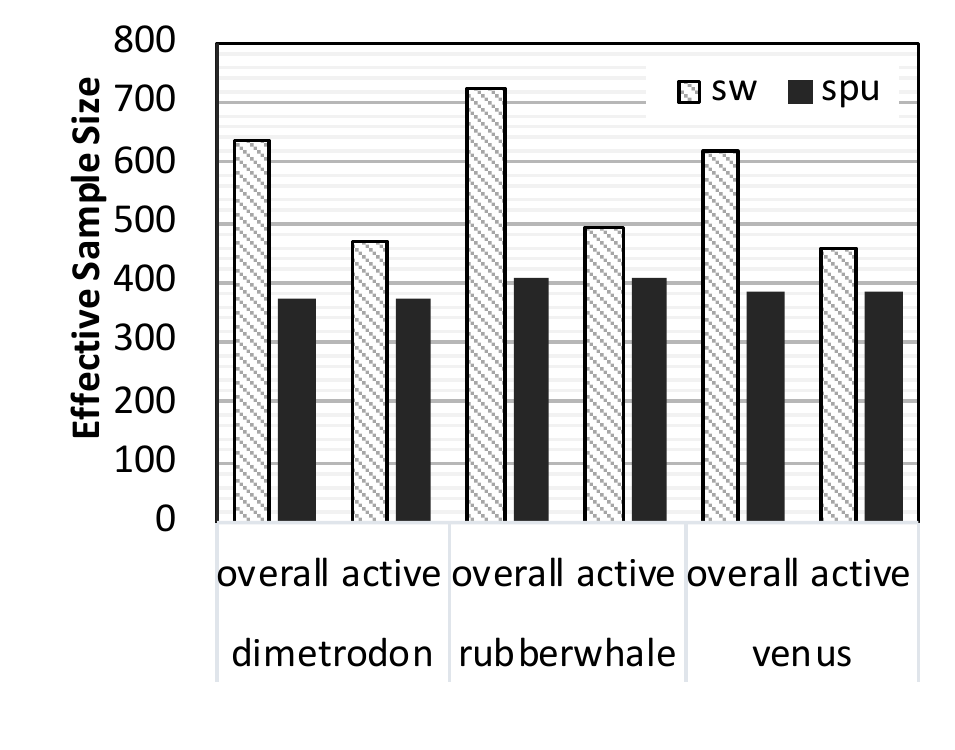}
    } 
	\caption {Mean overall and active ESS (higher is better) in each dataset. }
\label{fig:case_ess}
\end{figure}

\subsubsection{Sampling Quality}
We analyze ESS on SPU compared with the FP64 software by collecting the last 1,000 iterations of MCMC runs in the two applications. We evaluate the ESS per random variable and report the arithmetic mean. Fig. \ref{fig:ess_per_rv} shows an example ESS per random variable in stereo vision \textit{teddy} dataset. 
Red regions indicate the random variables have zero variance, and thus ESS cannot be calculated. Due to truncating small probabilities to zero, more random variables in the SPU have zero variance than in the software. We consider a random variable with zero variance inactive. 
The percentage of inactive random variables with respect to the total (a.k.a. inactive percentage) in three stereo vision datasets are 26.9\% for \textit{art}, 44.6\% for \textit{poster}, and 26.2\% for \textit{teddy} in the SPU, compared with 0.3\% for \textit{art}, 4.1\% for \textit{poster}, and 1.4\% for \textit{teddy} in the software. Motion estimation exhibits similar inactive percentages. Zero variance means the probability of a possible label is large enough that all random samples pick the same label, which can indicate convergence. The variance of corresponding inactive random variables in software are consistently small, indicating the random variable are likely to consistently pick the same label as well---a concentrated distribution.  Therefore, high inactive percentage does not necessarily imply bad result quality. 
%Both zero variances and high ESSs indicate convergence. 

Fig. \ref{fig:case_ess} shows the ESS arithmetic mean for a single MCMC run per dataset. 
We verify that different runs have small ESS difference ($<6$ in stereo vision). The mean ``overall" ESS eliminates the random variables with zero variance in software and hardware, respectively. Fig. \ref{fig:ess_per_rv} reveals that the inactive regions in the SPU (red) correspond to the regions with high ESS in software due to small but non-zero variance (yellow), and thus overall ESS is biased toward  software. Therefore, we also report the mean ``active" ESS which only includes the regions with non-zero variance in both software and the SPU, where ESS is more meaningful. As a consequence, the active ESS eliminates the regions with small variance in the software, which can potentially benefit the SPU. The importance of these small variances needs to be evaluated and we are actively looking for methods to account for these regions. The software has 1.1-1.4$\times$ higher active ESS than the SPU in stereo vision and around 1.2$\times$ in motion estimation. This implies the SPU needs to run 1.1-1.4$\times$ iterations to reach the same active ESS as the software. 

\subsubsection{Convergence Diagnostic}

\begin{figure}
\centering
	\captionsetup[subfigure]{width=0.20\textwidth}
	\subfloat[Stereo vision]{
		\includegraphics[width=0.27\textwidth,keepaspectratio,trim=5mm 0mm 0mm 0mm]{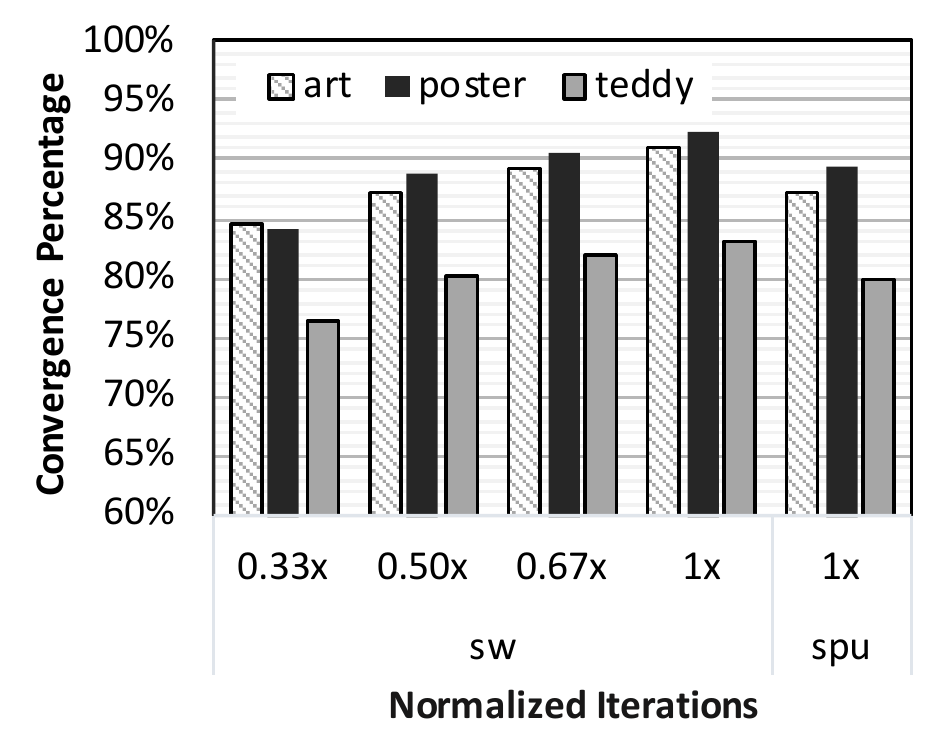}
    }
    \subfloat[Motion estimation]{
        \label{fig:case_cp_me}
    	\includegraphics[width=0.19\textwidth,keepaspectratio,trim=5mm 0mm 0mm 0mm]{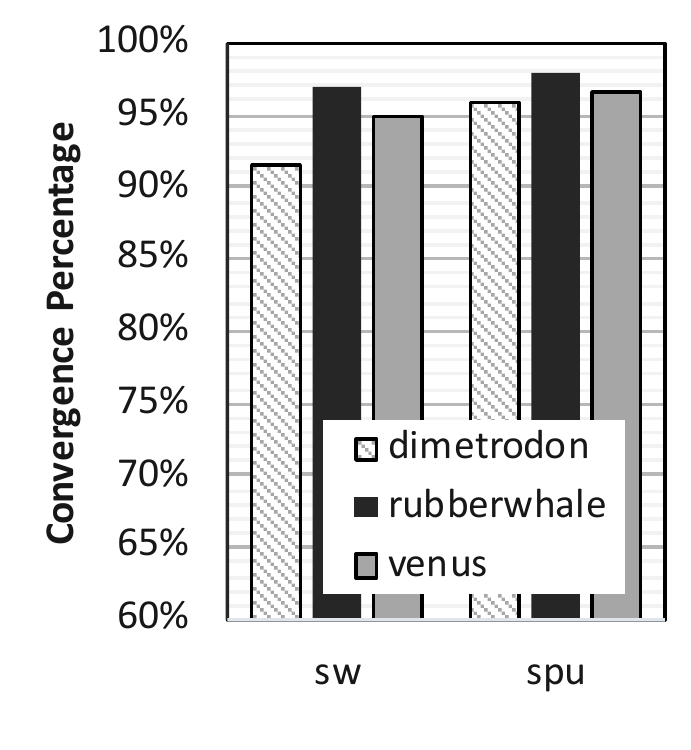}
    } 
	\caption{Convergence percentage (higher is better) results. The software and the SPU results in motion estimation are for the same number of iterations.}
\label{fig:case_cp}
\end{figure}

We evaluate the convergence diagnostic of SPU using the proposed convergence percentage metric. Each convergence percentage value is collected from 10 runs per dataset. Each run forfeits the first half of iterations as the burn-in period and only keeps the second half, as proposed by Gelman et al. \cite{gelman1992inference}. Recall a random variable is considered converged if $\hat{R}<1.1$ or both within-chain variance $W$ and between-chain variance $B$ are zero. The 2D motion vector is considered as two random variables in motion estimation. Fig. \ref{fig:case_cp} shows the results. 
The number of iterations is normalized with respect to SPU runs in stereo vision and are the same in motion estimation.  Overall, convergence percentage is high in both the software and the SPU: more than 80\% of random variables in stereo vision and more than 90\% in motion estimation. More than 99.5\% of random variables with $W=0$ in SPU are converged. In stereo vision, the SPU reaches the same or better convergence percentage than software with 2$\times$ iterations.
This indicates the SPU needs to be at least 2$\times$ faster in order to have a better overall performance in this application in terms of convergence percentage. Previous work \cite{Wang:isca2016,zhang2018isca} shows that a pipeline with the same data interface provides the speedups of at least 2.8-5.5$\times$ and up to 84$\times$. The SPU has higher convergence percentages than software in motion estimation, indicating the SPU converges faster in this application. Note that converging to a distribution faster does not necessarily lead to a better end-point result. The goodness of fit should be evaluated. 
%once converged. 

\subsubsection{Goodness of Fit}

\begin{figure}
\centering
	\captionsetup[subfigure]{width=0.20\textwidth}
	\subfloat[Stereo vision]{
		\includegraphics[width=0.23\textwidth,keepaspectratio,trim=13mm 10mm 10mm 10mm]{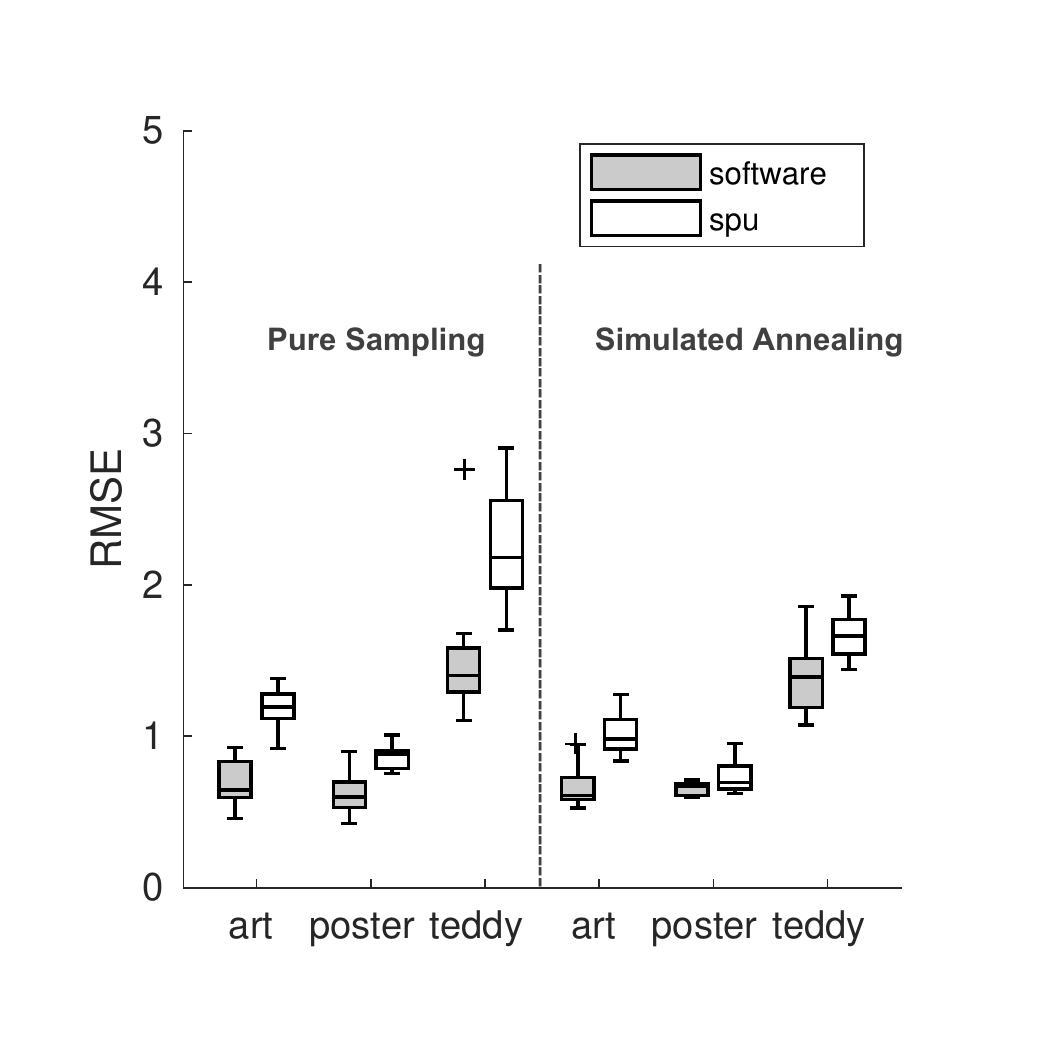}
    }
    \subfloat[Motion estimation]{
        \label{fig:case_rmse_me}
    	\includegraphics[width=0.23\textwidth,keepaspectratio,trim=15mm 10mm 10mm 10mm]{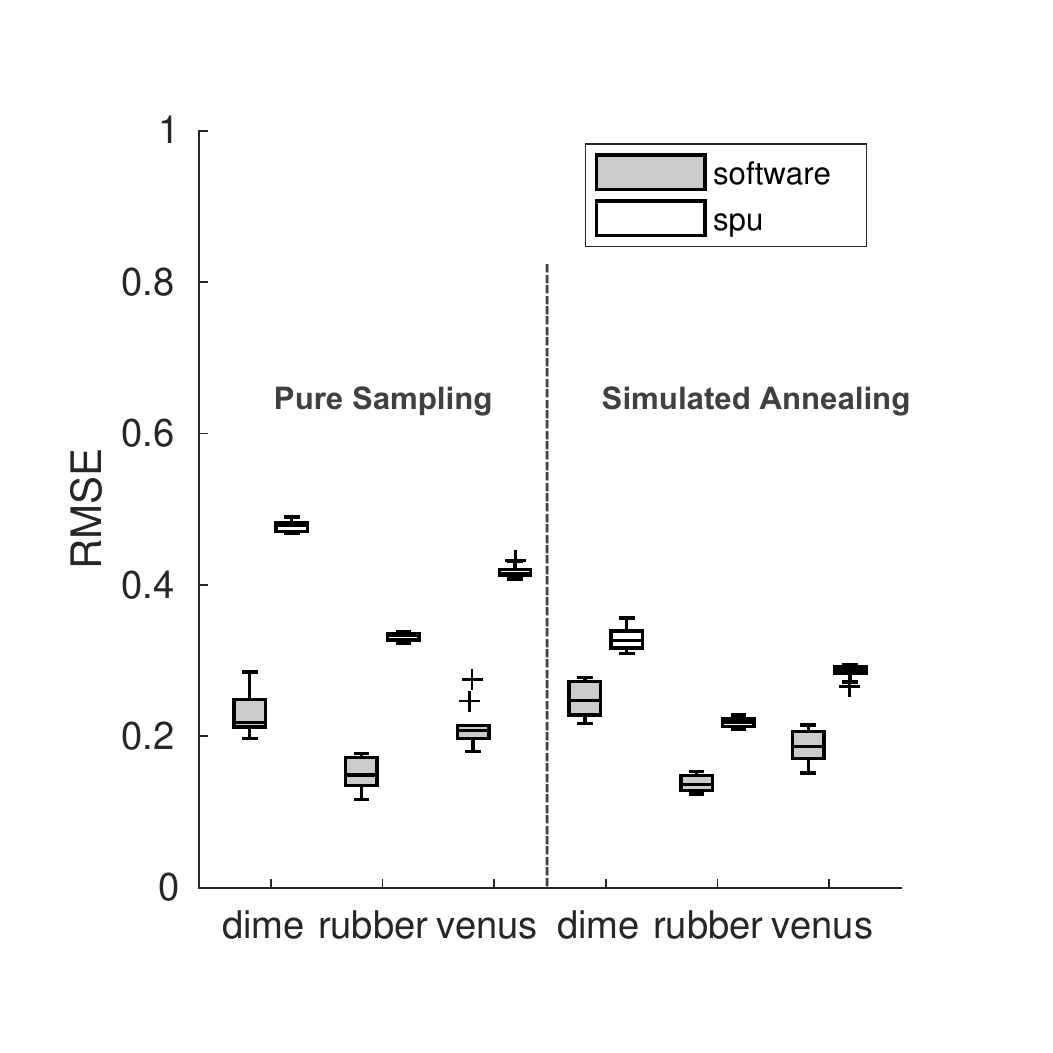}
    } 
	\caption{Root Mean Squared Error (lower is better). Note that scales are different in (a) and (b) due to application differences. }
\label{fig:case_rmse}
\end{figure}

\begin{figure}
\centering
	\captionsetup[subfigure]{width=0.20\textwidth}
	\subfloat[Stereo vision]{
		\includegraphics[width=0.24\textwidth,keepaspectratio,trim=10mm 10mm 10mm 10mm]{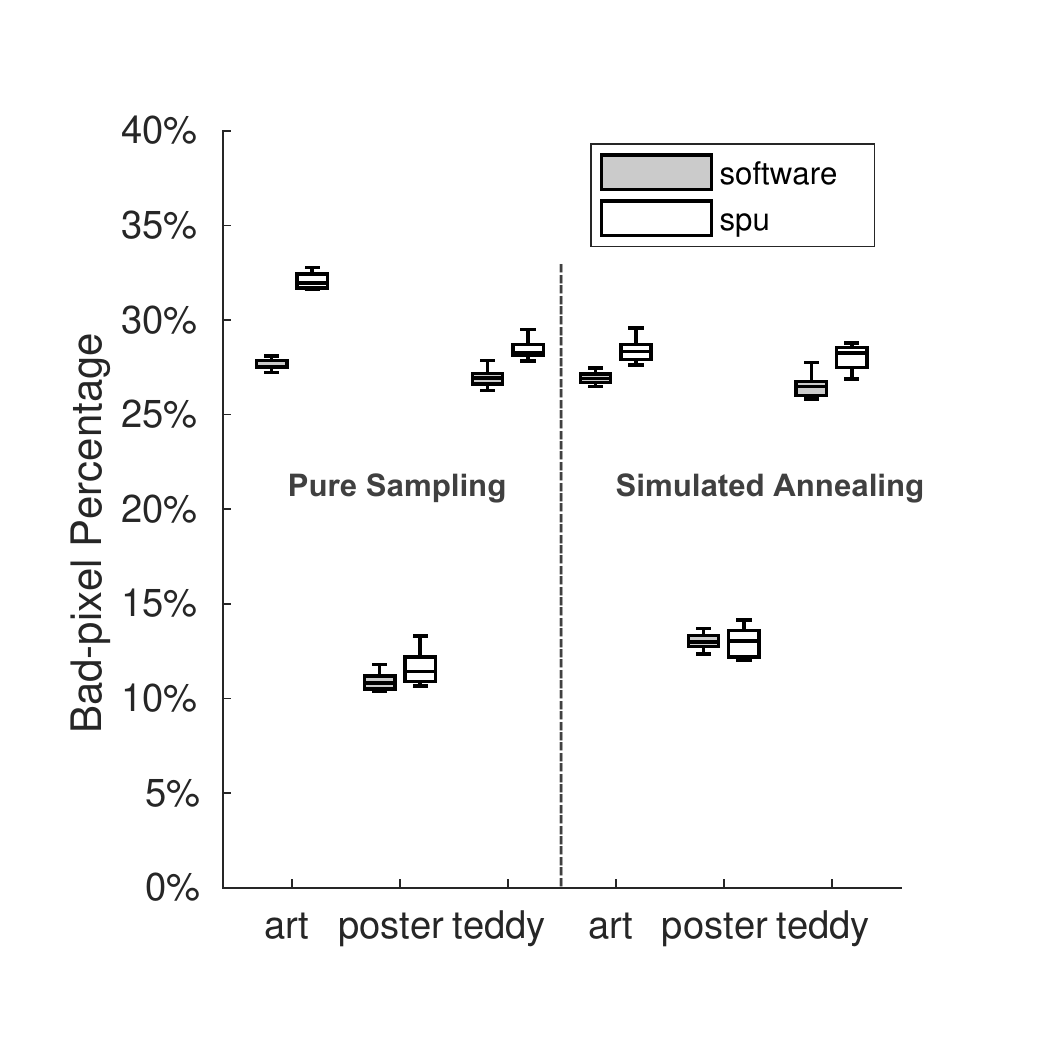}
    }
    \subfloat[Motion estimation]{
        \label{fig:case_epr_me}
    	\includegraphics[width=0.22\textwidth,keepaspectratio,trim=13mm 10mm 13mm 10mm]{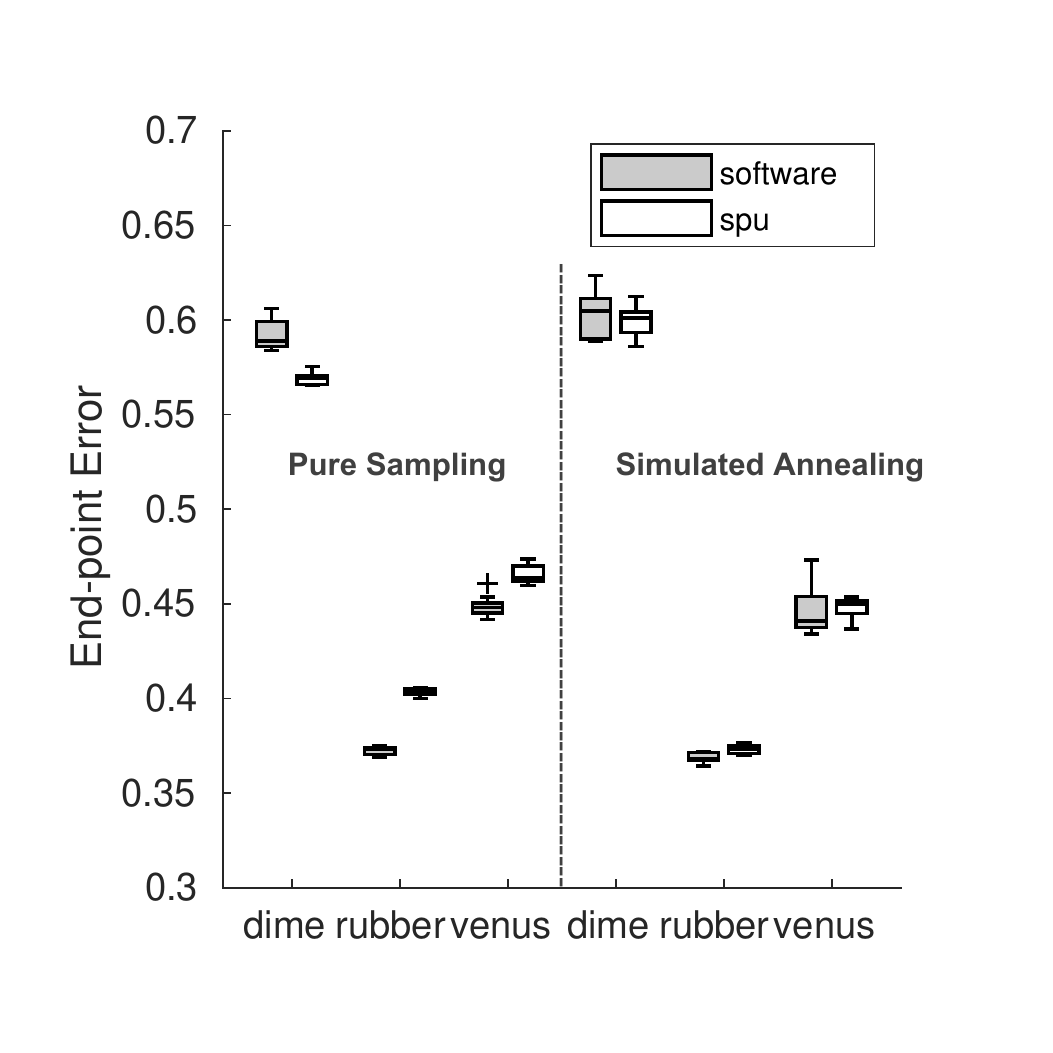}
    } 
	\caption{Application end-point result quality (lower is better). }
\label{fig:case_epr}
\end{figure}

The goodness of fit provides a similarity/difference measure on end-point results produced by the MCMC accelerator compared with results produced by the software. Fig. \ref{fig:case_rmse} shows the RMSE box plots of 10 MCMC runs per dataset compared with a reference result obtained by the mode of 10 software runs per dataset. Solid boxes show the range from 25th to 75th percentile. 
The horizontal lines inside each box are the medians of the data. The whiskers include the range of data that are not considered as outliers. We use 1.5$\times$ interquartile range as the rule to decide outliers, showed as pluses. Whiskers of the software and the SPU overlap in all stereo vision benchmarks, suggesting close results. RMSE results in motion estimation are visually different in Fig. \ref{fig:case_rmse_me}. However, these differences are small considering the small scale of y-axis. The software and the SPU produce closer results in simulated annealing optimization mode. 

Fig. \ref{fig:case_epr} shows the end-point result quality using ground truth data and application metrics. Most whiskers of the software and the SPU overlap except for \textit{art} in stereo vision and \textit{rubberwhale} in motion estimation, both of which are in sampling mode. In optimization mode, software and SPU whiskers overlap, indicating the difference in end-point result quality is very small. 
This is consistent with the single-run results in Tab. \ref{table:spu_prelim_results}. Note that no obvious differences between software and the SPU are visually observable in the stereo vision disparity maps and motion estimation flow maps. 

\begin{figure}
        \centering
  	\includegraphics[width=0.48\textwidth,keepaspectratio,trim=30mm 110mm 30mm 105mm]{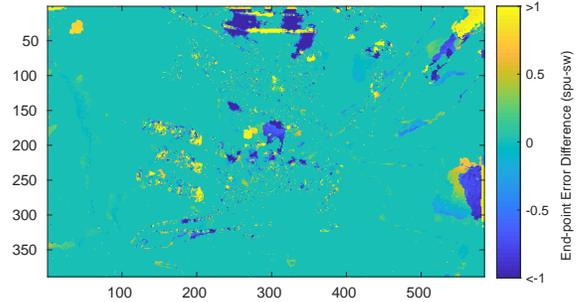} 
        \caption{\textit{Dimetrodon} end-point error difference ($spu-sw$) at pixel level. End-point error is 0.581 in software vs. 0.567 in SPU. }
        \label{fig:dime_epe_diff}
\end{figure}

It seems intuitive to assume that FP64 software should produce no worse results than hardware with limited precision, truncation, and a simplified RNG.
% given the higher bit-precision representation and dynamic range. 
%If this assumption always holds, higher SPU RMSE results compared with the software reference indicate worse application end-point results. 
We find this assumption holds in most, but not all, cases. We observe that for sampling , \textit{dimetrodon} has consistently lower end-point result error in the SPU than in the software (see Fig.~\ref{fig:case_epr_me}), which is consistent with convergence percentage results, shown in Fig. \ref{fig:case_cp_me}. However, Fig.~\ref{fig:case_rmse_me} shows that the SPU RMSEs are higher than FP64 software. 
To better understand this result, we examine per-pixel end-point error differences between the software reference and the SPU, as shown
in Fig.~\ref{fig:dime_epe_diff}. Blue regions correspond to lower end-point error in the SPU and yellow to lower end-point error in software. The figure suggests the software and the SPU have strengths in different regions, which potentially leads to a high RMSE compared to the software reference. 
%The RMSE compared with the software reference accumulates these differences and produces a higher RMSE than other software runs. 
This result indicates two insights: 1) software with higher precision does not necessarily produce better application end-point result quality, and 2) a higher RMSE compared to software does not always indicate worse application end-point result quality. Although bad pixel-percentage results are consistent with RMSE in stereo vision, the general link between the goodness of fit measure and the application end-point result quality needs to be further explored. 
This confirms collectively applying three pillars beyond end-point result is necessary to evaluate correctness.  

\begin{figure}
        \centering
  	\includegraphics[width=0.46\textwidth,keepaspectratio,trim=0mm 5mm 0mm 0mm]{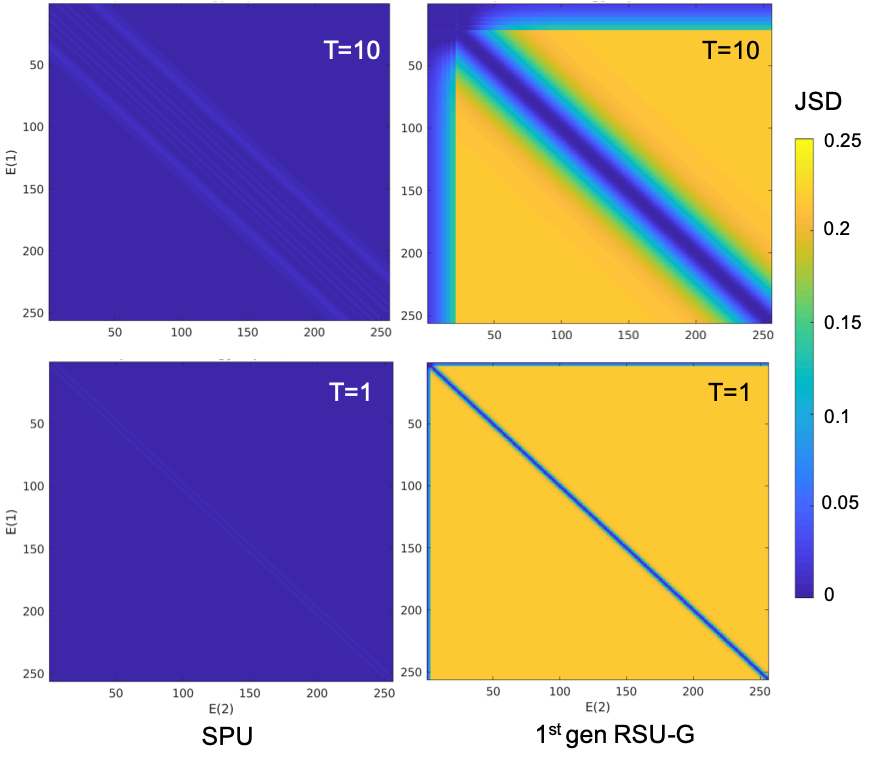} 
        \caption{Jensen-Shannon Divergence comparison between the SPU (left) and a previously proposed accelerator 1st-gen RSU-G (right).}
        \label{fig:jsd_example}
\end{figure}

We analyze the Jensen-Shannon Divergence of SPU relative to software with FP64 probability representation. Our goal is to provide insights about why hardware exhibits good or bad application end-point results and how it may perform with arbitrary input data.
We assume each random variable has a binary distribution in this analysis. By sweeping a wide range of possible energy inputs $E(i)$ from 0 to 255 in integer, corresponding to arbitrary data inputs, Fig. \ref{fig:jsd_example} plots JSD for two temperatures (1 and 10) and two different SPU microarchitectures: 1) the SPU described in Sec.~\ref{sec:SPU} and 2) an early design described by Wang et al.~\cite{Wang:isca2016}--called RSU-G---that was shown to lack sufficient precision and dynamic range~\cite{zhang2018isca}. These results clearly show the problems with the 1st generation RSU-G. The more recent SPU JSDs are negligible in most energy inputs (blue regions), whereas the 1st-gen RSU-G has high JSD for many inputs (yellow). 
%The worst-case SPU JSD of 0.012 occurs at $T=10$ when energy inputs have absolute difference of 34, and 0.007 at $T=1$ when absolute difference is 3. Input data that produces these energy differences have the worst-case divergence. 
The interpretation of these absolute JSD values are up to domain experts. 
The 1st-gen RSU-G JSD is greater than 0.2 for most energy inputs and becomes worse when temperature decreases, which explains the bad application result quality. A key difference between these two designs is dynamic scaling for energy values in the SPU, which is not present in the 1st-gen RSU-G.

% We also anlaysis the JSD on 1st generation RSU-G, the accelerator proposed previously \cite{Wang:isca2016}, which is claimed to have insufficient precision and dynamic ranges based on the application result quality \cite{zhang2018isca}. One of the major differences is 1st generation RSU-G does not dynamically scaling the energy values, lacking stage 2 in the SPU pipeline. Apparently, 

%% file: DesignSpace.tex
\section{Design Space Exploration: A Case Study}
\label{sec:dse}

The previous section shows that architectural optimizations might have negative influence on the statistical robustness, even though producing comparable end-point results to FP64 software. 
The question is \textit{can we achieve desirable end-point result quality and statistical robustness without the commensurate overhead of FP64?} 
To answer this question, we use the three pillars to explore the SPU design space.

\subsection{Design Trade-offs}
The SPU pipeline (Fig. \ref{fig:spu}) has several design parameters related to bit precision that potentially influence statistical robustness, including energy $E(i)$ and $E_s(i)$, scaled and truncated probability $p_{tr}(i)$, and RNG output bits.  We fix energy $E(i)$ and $E_s(i)$ at 8 bits based on previous work~\cite{mansinghka2014building,zhang2018isca}. 
The number of bits in $p_{tr}(i)$ considerably influences the size of the energy-to-probability converter and the discrete sampler. We evaluate three design points with 4-bit, 6-bit, and 8-bit $p_{tr}(i)$s. The influence of RNG output bits is small compared to $p_{tr}(i)$ and we find a 19-bit LFSR with 12-bit RNG outputs does not reduce the statistical robustness or result quality across all design points.

The SPU truncates all $p_{tr}(i)$s to the nearest $2^n$ values, called $2^n$ approximation \cite{zhang2018isca}, enabling efficient energy-to-probability conversion by comparing the boundaries of energy values. Without $2^n$ approximation, a double-buffered 256-entry LUT is required to store the $p_{tr}(i)$ values to achieve a stall-free design. We evaluate the statistical robustness of each scaled probability design point with and without $2^n$ approximations.
The above design parameters generally do not directly influence the SPU per-iteration performance assuming the same interface at the same clock frequency. However, a design with lower precision may take more iterations to converge. On the other hand, higher precision requires more area and power affecting the number of SPU units in systems with limited area/power budget. Thus we discuss resource usage in Sec.~\ref{sec:resource}. Detailed system-level architecture investigations are beyond the scope of this paper.

\subsection{Evaluating Design Parameters}

\begin{figure}
\centering
	\captionsetup[subfigure]{width=0.48\textwidth}
    \subfloat[Overall ESS]{
        \label{fig:dse_overall_ess}
    	\includegraphics[width=0.4\textwidth,keepaspectratio,trim=15mm 15mm 0mm 5mm]{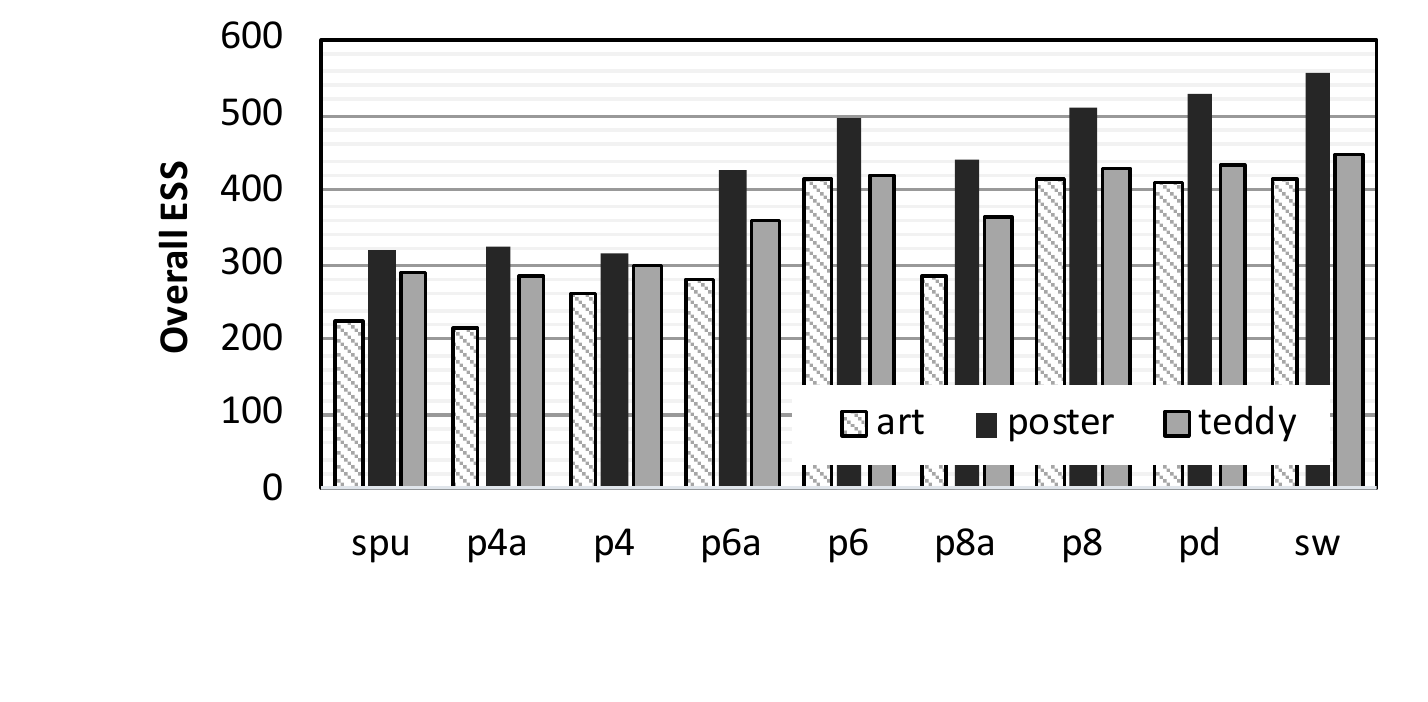}
    } 
    \\	
    \subfloat[Percentage of inactive random variables (inactive percentage).]{
        \label{fig:dse_inact_percentage}
    	\includegraphics[width=0.4\textwidth,keepaspectratio,trim=15mm 15mm 0mm 5mm]{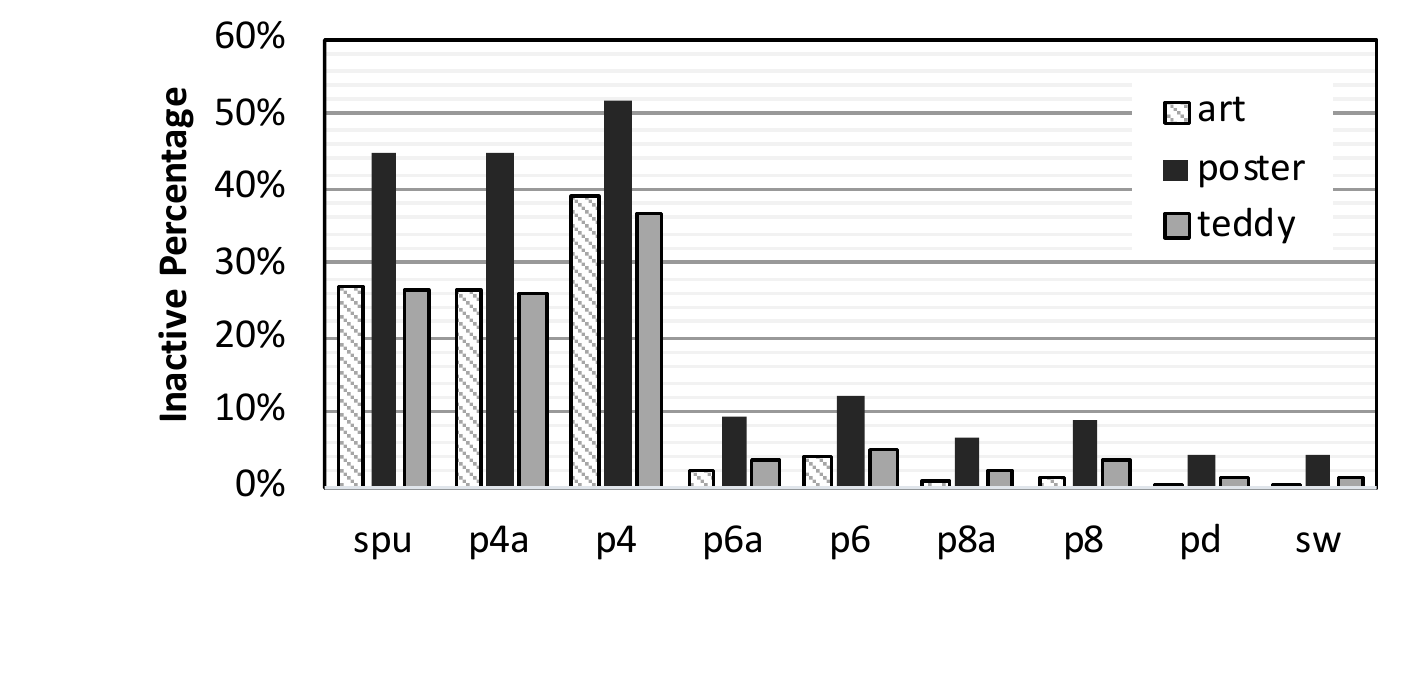}
    } 
    \\
	\subfloat[Active ESS in \textit{teddy}]{
	    \label{fig:dse_active_ess_teddy}
		\includegraphics[width=0.4\textwidth,keepaspectratio,trim=15mm 15mm 0mm 5mm]{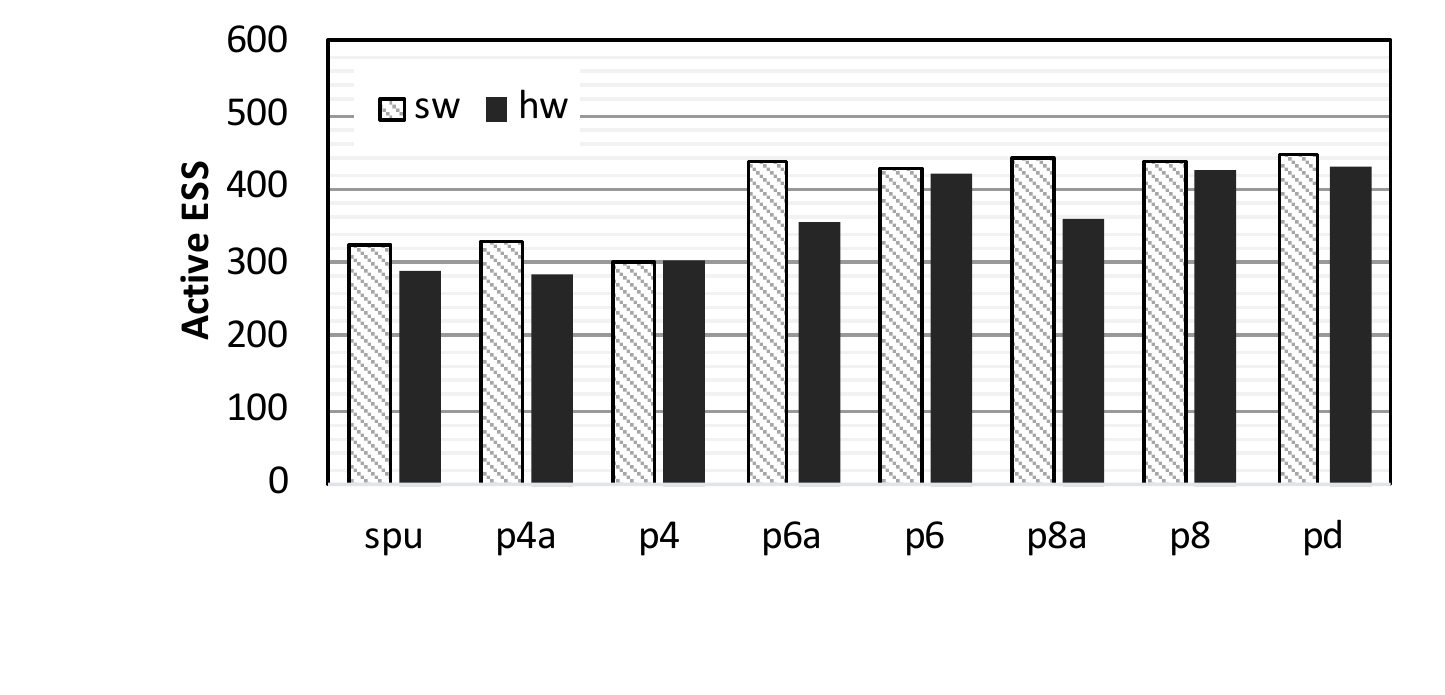}
    } 
	\caption{Stereo vision sampling quality in the design points.}
\label{fig:dse_ess}
\end{figure}

\begin{figure}
        \centering
  	\includegraphics[width=0.4\textwidth,keepaspectratio,,trim=15mm 15mm 0mm 5mm]{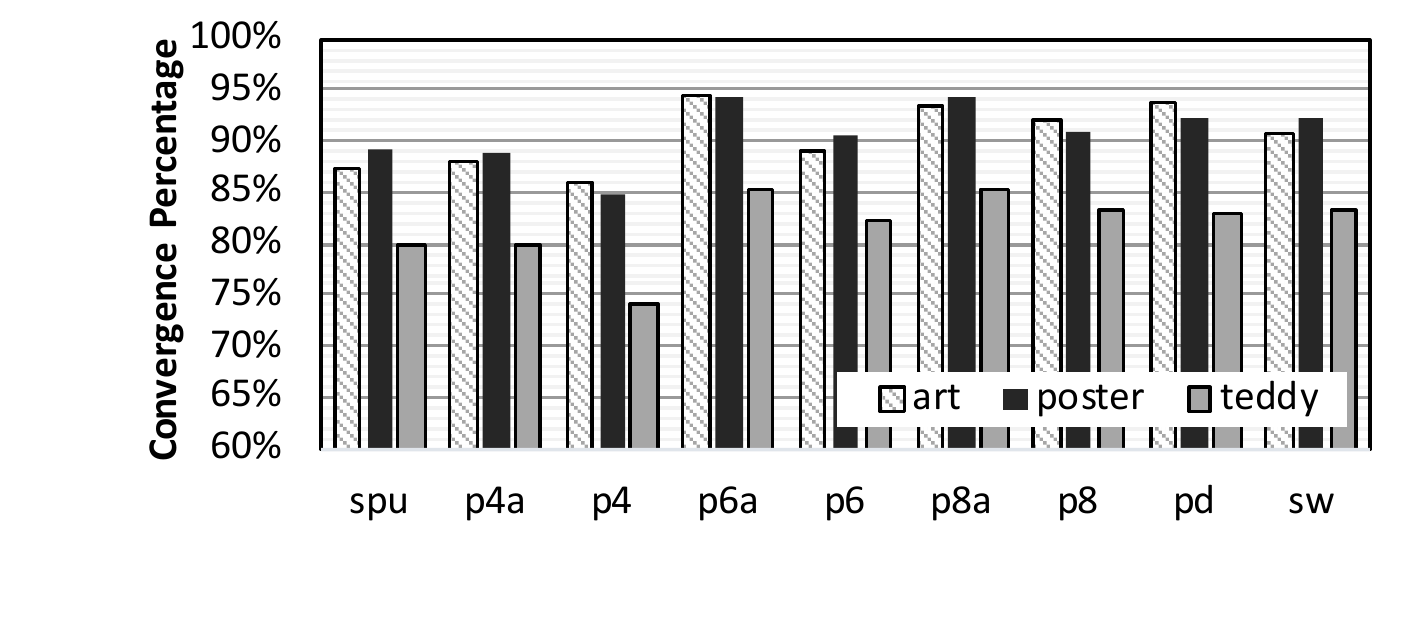} 
        \caption{Stereo vision convergence percentage in the design points.}
        \label{fig:dse_cp}
\end{figure}

\begin{figure}
\centering
	\captionsetup[subfigure]{width=0.48\textwidth}
	\subfloat[Pure sampling (sampling)]{
		\includegraphics[width=0.49\textwidth,keepaspectratio,trim=19mm 5mm 24mm 5mm]{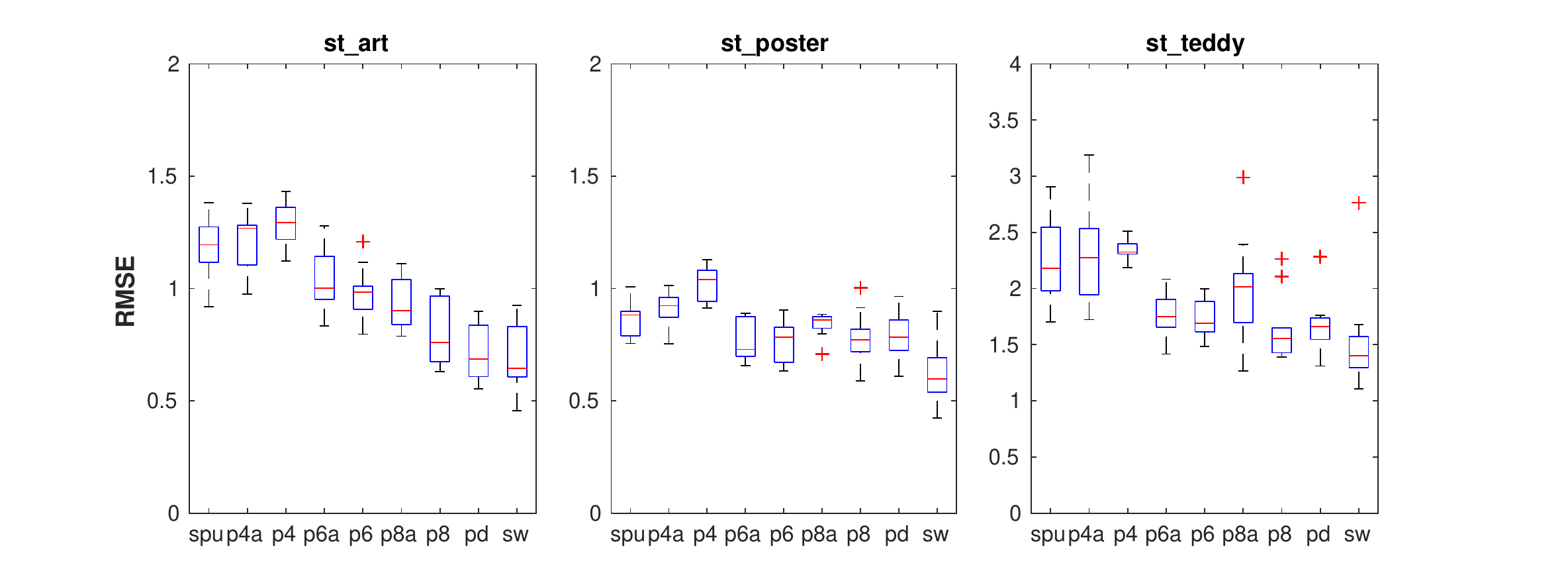}
    } 
    \\
    \subfloat[Simulated annealing (optimization)]{
    	\includegraphics[width=0.49\textwidth,keepaspectratio,trim=19mm 5mm 24mm 5mm]{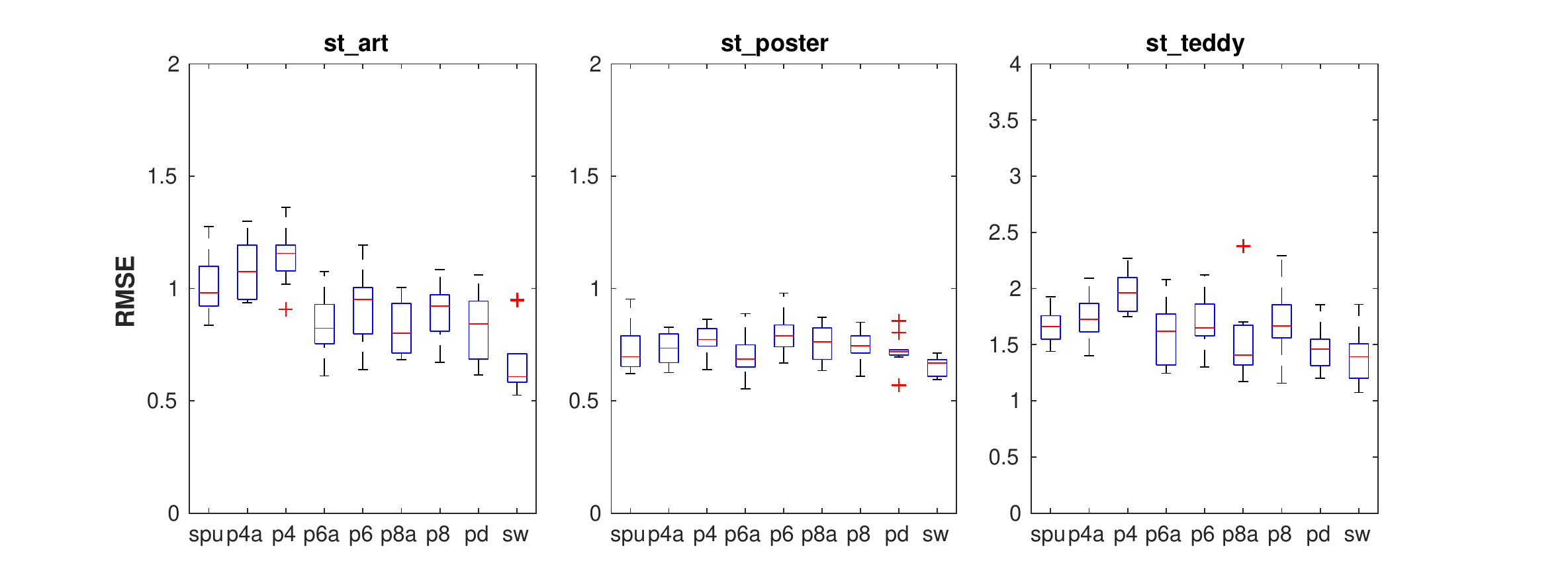}
    } 
	\caption{Stereo vision RMSE in the design points.}
\label{fig:dse_rmse}
\end{figure}

\begin{figure}
\centering
	\captionsetup[subfigure]{width=0.48\textwidth}
	\subfloat[Pure sampling (sampling)]{
		\includegraphics[width=0.49\textwidth,keepaspectratio,trim=19mm 5mm 24mm 5mm]{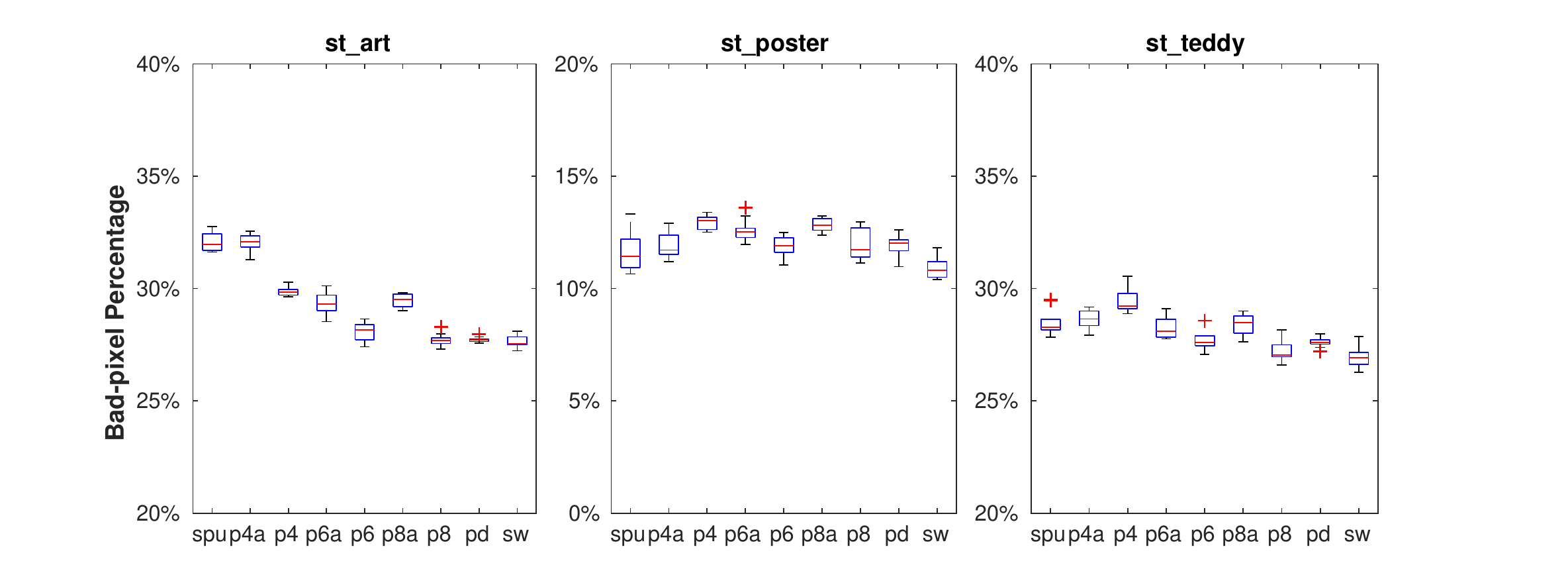}
    } 
    \\
    \subfloat[Simulated annealing (optimization)]{
    	\includegraphics[width=0.49\textwidth,keepaspectratio,trim=19mm 5mm 24mm 5mm]{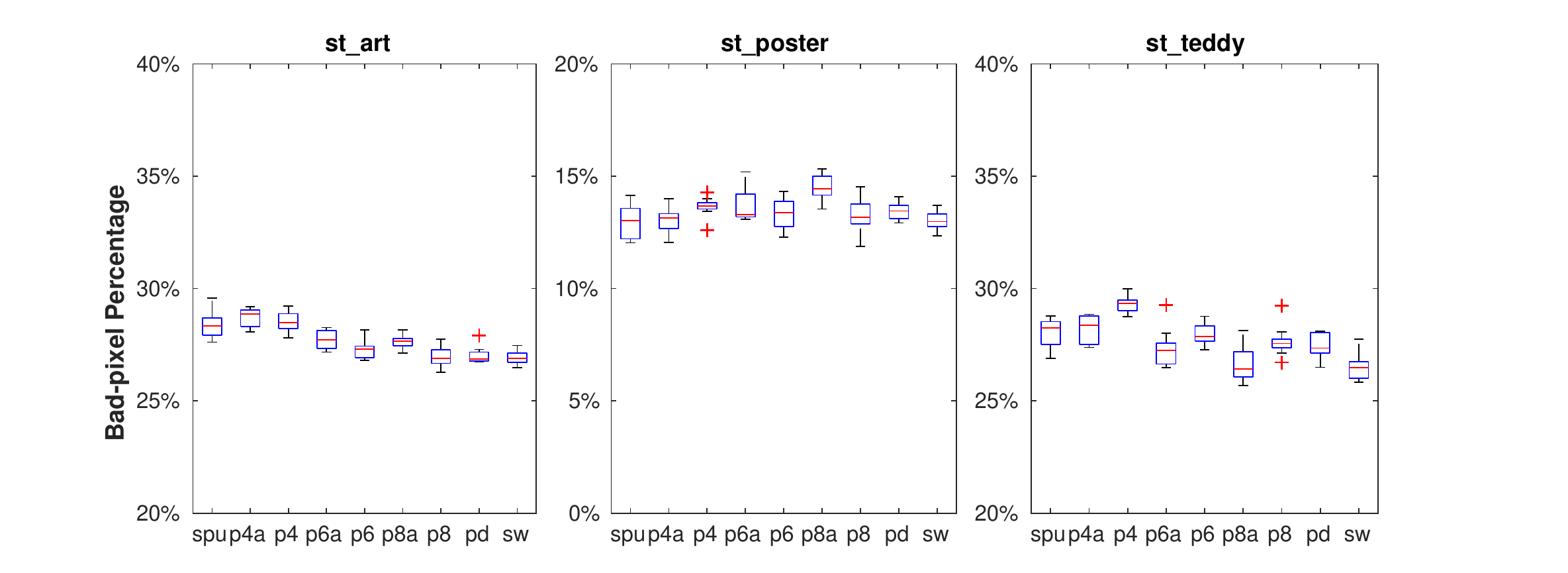}
    } 
	\caption{Stereo vision application end-point result quality in the design points.}
\label{fig:dse_epr}
\end{figure}

Figs. \ref{fig:dse_ess}-\ref{fig:dse_epr} show our design space results. For brevity, we show only stereo vision results and highlight motion estimation results where needed. Starting from the current SPU design (``spu" label in figures), we analyze the statistical robustness by gradually increasing the precision. First, we replace the 19-bit LFSR sampler with a FP64 Mersenne Twister sampler~\cite{matsumoto1998mersenne} while keeping the front-end pipelines unchanged (``p4a"). 
Next, we increase the bit width of $p_{tr}(i)$ to 6-bit (``p6a") and 8-bit (``p8a"), with $2^n$ approximation and then remove $2^n$ approximation corresponding to ``p4", ``p6", and ``p8" in the figures. We further evaluate an upper-bound design (``pd") that keeps front-end pipeline up to the scaled energy ($E_s(i)$) output unchanged, but has a FP64 back-end for probability conversion and discrete sampling.

\subsubsection{Sampling Quality} 
Fig. \ref{fig:dse_overall_ess} shows the overall ESS, which omits random variables with zero variance for each design, respectively. Recall this metric can create bias that benefits software for variables with small but non-zero variance. Overall ESSs increase when more bits are added, partly as a result of fewer random variables with zero variance. Recall the SPU truncates small scaled probabilities $p_{tr}(i)<1$ to zero. Adding more bits keeps more possible labels with small probabilities available to be sampled. 
Fig. \ref{fig:dse_inact_percentage} indicates inactive percentage drops significantly when increasing $p_{tr}(i)$ bit size from 4 to 6. Interestingly, $2^n$ approximation helps reduce the inactive percentage under the same bit precision, but decreases ESS for 6-bit and 8-bit designs. 
Fig. \ref{fig:dse_active_ess_teddy} shows the active ESS for the \textit{teddy} dataset. Recall active ESS masks out the random variables inactive in either software or the SPU.  With $2^n$ approximation, increasing bit precision does not close the gap in active ESS with the software. 
Without $2^n$ approximation, 6 or 8-bit $p_{tr}(i)$ have comparable overall and active ESS to software. As expected, increasing bit precision decreases the difference between overall and active ESS due to fewer inactive variables. Results for motion estimation (omitted) are similar. 

\subsubsection{Convergence Diagnostic} Fig. \ref{fig:dse_cp} shows the convergence percentage increases with the increasing bit precision. In contrast to ESS, $2^n$ approximation improves the convergence percentage under the same bit precision. Hardware with 6-bit and 8-bit $p_{tr}(i)$ with and without $2^n$ approximation produces comparable convergence percentage to software. All designs except ``p4" produce
the same or higher convergence percentage for motion estimation (results not shown). 

\subsubsection{Goodness of Fit} Fig. \ref{fig:dse_rmse} shows RMSE results compared with software reference results. Observable lower RMSEs can be found in stereo vision \textit{art} when increasing the bit precision from 4 to 6. Differences of RMSEs are hard to notice when further increasing the precision given whiskers largely overlap in most datasets. 
Application end-point results in Fig. \ref{fig:dse_epr} exhibit the same trends. All designs produce comparable result quality to the software in simulated annealing (optimization), consistent with Tab. \ref{table:spu_prelim_results}. 
We highlight the following results for motion estimation (not shown): 1) the design parameters have negligible influence on application end-point result quality (end-point error) except ``p4" in a couple of cases, which performs observably worse, 2) all designs except ``p4" produce better end-point error than software for \textit{dimetrodon} with sampling, 3) all designs produce slightly worse end-point error than the software for \textit{rubberwhale} with sampling, 4) gaps exist between the software and all hardware designs including ``pd" for RMSE, but not in end-point error, which confirms the importance of using all three pillars.
Overall, optimization is more robust than sampling at producing good result quality across various designs. For both modes, increasing the scaled probability to 6 bits produces comparable goodness of fit results to the software.

\subsection{Resource Usage}
\label{sec:resource}
\subsubsection{FPGA Resource Usage}

\begin{scriptsize}
\begin{table}
  \centering
  \caption{Resource usage and performance of various SPU implementations on Arria 10 FPGA.}
  \label{table:FPGA_Results}
  \begin{tabular}{|l|l|l|l|}
    \hline
    \textbf{Parameter} & \textbf{Verilog} & \textbf{HLS-int} & \textbf{HLS-fp}\\
    \hline
    \hline
    Frequency (MHz) & 374 & 369 & 320\\
    \hline
    ALMs & 321 & 1,189 & 4,407\\
    \hline
    Registers & 680 & 2,551 & 7,932\\
    \hline
    Memory Bits & 1,472 & 10,688 & 49,536\\
    \hline
    DSPs & 4 & 10 & 25\\
    \hline
    Initiation Interval (Cycles) & 1 & 1 & 3\\
    \hline
  \end{tabular}
\end{table}
\end{scriptsize}

We evaluate three different implementations of the SPU on an Intel Arria 10 FPGA \cite{Quartus}: 1) an optimized hand-written Verilog implementation with 4-bit scaled probability, 2) a High-Level Synthesis (HLS) implementation (HLS-int) that matches the hand-written Verilog (but using HLS basic integer data-types), and 3) an HLS implementation with a 32-bit floating-point back-end after energy computation (HLS-fp), eliminating the energy scaling stage. 
HLS-fp is developed in order to assess the option of directly using 32-bit floating-point representation inside the SPU for probability conversion and sampling. 
Tab. \ref{table:FPGA_Results} shows the synthesis results. Despite higher resource requirement, HLS-int is close to the Verilog implementation in terms of performance. The resource usage can be further decreased by using reduced-precision integers. 
HLS-fp consumes significantly more resources compared to HLS-int 
and most importantly performs remarkably worse due to its lower operating frequency (369 MHz vs. 320MHz) and lower throughput (1 vs. 3 initiation interval cycles) caused by the floating-point addition \cite{FloatingPoint}. Clearly, naively implementing the SPU in 32-bit floating-point consumes too much resources and significantly reduces the performance benefits. A human-designed architecture is needed to improve efficiency.

\begin{scriptsize}
\begin{table}
  \centering
  \caption{Area and Power Analysis in ASIC}
  \label{table:asic_results}
  \begin{tabular}{|l|l|l|l|l|l|}
    \hline
    \textbf{Design} & \textbf{Area ($\mu m^2$)}& \textbf{Power (mW)} & \textbf{Design} & \textbf{Area} & \textbf{Power} \\
    \hline
    \hline
    spu & 1957 & 2.17 & p4 & 2112 & 2.21\\
    \hline
    p6a & 2134 & 2.31 & p6 & 2356 & 2.38\\
    \hline
    p8a & 2309 & 2.46 & p8 & 2599 & 2.54\\
    \hline
  \end{tabular}
\end{table}
\end{scriptsize}

\subsubsection{ASIC}

We estimate the ASIC area/power for various design points. 
Circuits elements written in Chisel are synthesized in a predictive 15nm library \cite{Martins2015} using Synopsys Design Compiler.  Memory elements (FIFOs and LUTs) are estimated using Cacti 7 \cite{balasubramonian2017cacti} in 22nm technology, the smallest supported technology. 
The designs are verified in stereo vision \textit{art}. 
Tab. \ref{table:asic_results} summarizes the results. Total area/power numbers are the sum of 15nm circuitry and 22nm memory elements. Power is estimated at 1GHz. Since Cacti requires widths in multiples of bytes, we estimate a double-buffered 2$\times$256-byte LUT (537 $um^2$ and 0.32 $mW$) and a 64-byte FIFO (215 $um^2$ and 0.18 $mW$) with 8-bit port, and linearly scale them to target widths of 4 and 6 bits. Overall, modest overheads are introduced when slightly increasing the bit precision from 4 to 8. All designs can run up to 3.3GHz, bounded by the SPU energy computation stage.  
Increasing the SPU $p_{tr}(i)$ from 4-bit to 6-bit precision while keeping the $2^n$ approximation (``p6a") incurs 1.09$\times$ area and 1.07$\times$ power overheads, but has considerably better statistical robustness. Removing $2^n$ approximation (``p6a") adds double-buffered LUTs for energy-to-probability conversion, thus incurs 1.20$\times$ area and 1.10$\times$ power overheads. Despite a 10\% difference in area, we advocate the 6-bit designs without $2^n$ approximation in an ASIC for better sampling quality if area is not a major concern. The benefit from further increasing the bit-precision is marginal based on the previous analysis.

%% file: LimFuture.tex
\section{Limitations and Future Work}
\label{sec:future}

The proposal of three pillars is an important starting point towards quantifying the statistical robustness of probabilistic accelerators. This work selects the most popular metrics and estimation approaches from many within each pillar. The analysis of other metrics and methods (e.g., MCMC standard error \cite{flegal2008markov}) might help identify limitations of selected metrics. 
The challenges of naively applying existing metrics motivate us to propose modified processes and a new metric for reporting scalar measures for sampling quality and convergence diagnostic. 
Our proposals are conceptually straightforward, but could benefit from domain experts developing metrics with stronger theoretical foundations. 
Ideally, formally proving bounds on the metrics for an accelerator could provide guarantees on statistical robustness, but is extremely difficult or impossible due to many hardware approximations techniques (e.g., truncation to zero). Additionally the adequateness of rule-of-thumb $\hat{R}<1.1$ to determine convergence is under debate \cite{vats2018revisiting}. 
%Although the period is notably short compared with other popular softwrae RNGs, 
%The simple 19-bit LFSR does not reduce the statistical robustness of the hardware in the tested application. Understanding the influence of RNGs on the probabilistic algorithms is very challenging and needs continual efforts \cite{click2011quality}.
Applying the three pillars on other accelerators, applications, and models is our future work. Our proposed processes and metric apply to other accelerators and applications that cannot directly use the existing methods due to high dimensionality and potential random variables with zero variance. %Applications with continuous random variables and low dimensionality can apply the existing methods. 
The effects of hardware approximations are unknown for applications that require information from variables with very low variance, such as rare event simulation.

% Other metrics
% Although 
% 19-bit LFSR, as long as other PRNG, is predictable. System needs to prevent them for being controlled by malicious attacks.  
% analytical approaches. Analytically quantifying a complex system is challenging. (Why? The dynamic scaling. )
% Rare event is challenging use the hardware, we will lose the tail. 
% Rare event simulation
% Some recent work has questioned whether threshold of complete $\hat{R}<1.1$ is too high \cite{vats2018revisiting}, our method applies to any $\hat{R}$ thresholds. 

% training on hardware. 

% Future work include but not limited to: other MCMC accelerators, other applications. Methods with We invite domain experts to evaluate and provide insights on our proposed method. 

%% file: RelatedWork.tex
\section{Related Work}
\label{sec:related}
%\mike{remind me if I miss some works. Move this to appropriate places if needed.}

% Provide guarantees in neural networks. 
% Vikash works on QQ plots and KL-divergence. 
% UCL works custom precision,  . 
Sec. \ref{sec:spec_acc} discussed various specialized accelerators for probabilistic algorithms. Other accelerators exists for deterministic Bayesian Inference \cite{Lin2010hbc,hurkat2019vip}. A benchmark for Bayesian Inference models is proposed for performance evaluation \cite{wang2019demystifying}. 
Previous work addresses some statistical metrics for MCMC accelerators. Mansinghka et al. \cite{mansinghka2014building} evaluates data input precision using KL-divergence and QQ plots. Liu et al. \cite{liu2015} argues using ESS/second as a performance metric for MCMC samplers. 
Multiple goodness of fit statistical tests exist \cite{gretton2012kernel,saad2019family}. These metrics all belong to one of three pillars proposed in this work and we argue all three pillars are needed to fully characterize the statistical robustness of an MCMC accelerator. 
Theoretical studies provide error bounds for MCMC with algorithmic approximation techniques given mathematical assumptions \cite{johndrow2015optimal,ge2018simulated}. An analytical tool for quantization error is proposed to help hardware design decision \cite{linderman2010numerical}, but does not address the statistical robustness. 
Analytical and empirical studies have been done on evaluating limited precision in neural networks \cite{sakr17guarantee,FinitePrecisionMLP,BinaryConnect,DeepLearningLimitedPrecision}. 

%Attempt in neural networks. 
%There have been both analytical and empirical studies on the effects of limited precision computations on the output of neural networks. Holt and Hwang presented an analysis of the effects of low precision fixed-point arithmetic on a multi-layer perceptron \cite{FinitePrecisionMLP}. More recently, Courbariaux et al. proposed BinaryConnect \cite{BinaryConnect} which represents network weights using only two values. Gupta et al. studied the effect of limited precision data representation and computation on neural network training, and designed a hardware accelerator using 16-bit fixed-point numbers \cite{DeepLearningLimitedPrecision}.

%% file: Conclusion.tex
\section{Conclusion}

Domain-specific accelerators require correctness evaluation. In probabilistic algorithms, statistical robustness is an important aspect of correctness defined by domain experts. Current methodologies often omit statistical robustness and thus lack a comprehensive definition of correctness. 
This work takes a first step toward defining metrics and a framework for evaluating correctness of probabilistic accelerators beyond application end-point result quality.
We propose three pillars of statistical robustness: 1) sampling quality, 2) convergence diagnostic, and 3) goodness of fit. 
We apply the three pillars on an existing hardware accelerator and surface design issues that cannot be revealed by only using application end-point result quality. The three pillars guide the design space exploration and achieve considerable improvements in the statistical robustness by slightly increasing the bit precision. 
We believe our framework can help architects to design robust probabilistic hardware. 

\label{sec:conclusion}